\documentclass[submission, Phys]{SciPost}

\hypersetup{
    colorlinks,
    linkcolor={red!50!black},
    citecolor={blue!50!black},
    urlcolor={blue!80!black}
}

\usepackage[bitstream-charter]{mathdesign}
\usepackage{braket}
\usepackage{xcolor}
\usepackage{soul}
\DeclareSymbolFont{usualmathcal}{OMS}{cmsy}{m}{n}
\DeclareSymbolFontAlphabet{\mathcal}{usualmathcal}

\DeclareMathOperator{\sgn}{{\rm sgn}}
\newcommand{\ii}{{\rm i}}
\newcommand{\commutator}[2]{\left[#1, #2\right]}

\DeclareMathOperator{\sign}{sign}

\begin{document}

\begin{center}{\Large \textbf{
Entanglement R\'enyi Entropies from Ballistic Fluctuation Theory: 
the free fermionic case\\
}}\end{center}

\begin{center}
Giuseppe Del Vecchio Del Vecchio\footnote{{\small \sf giuseppe.del\textunderscore vecchio\textunderscore del\textunderscore vecchio@kcl.ac.uk}},
Benjamin Doyon\footnote{{\small \sf paola.ruggiero@kcl.ac.uk}},
Paola Ruggiero\footnote{{\small \sf paola.ruggiero@kcl.ac.uk}}
\end{center}

\begin{center}
Department of Mathematics, King’s College London, Strand, London WC2R 2LS, UK
\\
\end{center}

\newcommand{\gdv}[1]{\textcolor{orange}{Giuseppe : #1}}

\begin{center}
\today
\end{center}


\section*{Abstract}
{\bf
The large-scale behaviour of entanglement entropy in finite-density states, in and out of equilibrium, can be understood using the physical picture of particle pairs. However, the full theoretical origin of this picture is not fully established yet. In this work, we clarify this picture by investigating entanglement entropy using its connection with the large-deviation theory for thermodynamic and hydrodynamic fluctuations. We apply the universal framework of Ballistic Fluctuation Theory (BFT), based the Euler hydrodynamics of the model, to correlation functions of \emph{branch-point twist fields}, the starting point for computing R\'enyi entanglement entropies within the replica approach. Focusing on free fermionic systems in order to illustrate the ideas, we show that both the equilibrium behavior and the dynamics of R\'enyi entanglement entropies can be fully derived from the BFT. In particular, we emphasise that long-range correlations develop after quantum quenches, and accounting for these explain the structure of the entanglement growth. 
We further show that this growth is related to fluctuations of charge transport, generalising to quantum quenches the relation between charge fluctuations and entanglement observed earlier. The general ideas we introduce suggest that the large-scale behaviour of entanglement has its origin within hydrodynamic fluctuations.
}

\vspace{10pt}
\noindent\rule{\textwidth}{1pt}
\tableofcontents\thispagestyle{fancy}
\noindent\rule{\textwidth}{1pt}
\vspace{10pt}

\section{Introduction}
\label{sec:intro}

The understanding of entanglement in quantum many-body systems received a considerable boost in the
last decades, with the introduction and characterization of many different quantities which ‘‘measure'' the amount of entanglement in a given quantum state~\cite{amico2008entanglement,calabrese2009entanglement,eisert2010colloquium,laflorencie2016quantum}.
An important set of such measures are the so-called
entanglement R\'enyi entropies. Given a quantum system described by a density
matrix $\rho$ and a subsystem $A$ of the total system, with $\bar{A}$
denoting its complement, consider the associated reduced density matrix $\rho_{A}=\text{tr}_{\bar{A}}\rho$. Then, for any $\alpha\in\mathbb{R}^{+}$,
the $\alpha$-R\'enyi entropy is defined as 
\begin{equation}
S_{\alpha}=\frac{1}{1-\alpha}\log\text{\text{tr}\ensuremath{\rho_{A}^{\alpha}}}.\label{eq:renyi_def}
\end{equation}
They are good entanglement measures for all pure quantum
states, i.e.~states of the form $\rho=|\Psi\rangle\langle\Psi|$. They fully characterise the entanglement spectrum, and an important
property is that in the limit $\alpha\to1$ they reduce to the famous
entanglement Von Neumann entropy 
\begin{equation} 
S=-\text{tr}\left(\rho_{A}\log\rho_{A}\right).\label{eq:entent_def}
\end{equation}

In the context of one-dimensional systems, which is the focus of this paper, several exact results are available for such quantities. For example, at equilibrium, R\'enyi and entanglement entropies or their asymptotic behaviours can be obtained in the ground state states of critical~\cite{calabrese2009conformal}, gapped~\cite{calabrese2004entanglement} and more general integrable~\cite{cardy2008form} field theories, as well as beyond integrability \cite{PhysRevLett.102.031602} (note that for free theories results were first obtained in~\cite{casini2005entanglement}).
In the case of critical systems described by a conformal field theory (CFT), such results are easily generalized to finite temperature states (i.e., Gibbs ensembles)~\cite{calabrese2009conformal}, and also results for
generic thermodynamic macrostates  (i.e., generalized Gibbs ensembles~\cite{rigol2007relaxation}) have been obtained~\cite{alba2017quench,alba2017renyi,mestyan2018renyi} in the context of integrable models relying on (thermodynamic) Bethe ansatz~\cite{1931_Bethe_ZP_71} methods.

When moving to out-of-equilibrium scenarios, the situation is more complicated and available results are mainly qualitative or in the form of conjecture (an exception, however, is the exact result in~\cite{fagotti2008evolution}). For example, an imaginary time path-integral formulation, together with conformal invariance, has been used for a qualitative understanding of the ubiquitous linear growth of entanglement~\cite{calabrese2005evolution} observed after quantum quenches~\cite{calabrese2007quantum,calabrese2006time}. 
Moreover, the dynamics of the entanglement entropy~\eqref{eq:entent_def} for a generic integrable system was understood in terms of a semiclassical ``quasiparticle picture'' (whose original version was proposed in \cite{calabrese2005evolution}), complemented with the Bethe ansatz knowledge of the stationary state attained at late times, as conjectured in \cite{alba2017entanglement} (see also \cite{alba2018entanglementlong}). These results have been extensively verified numerically (see, e.g.,~\cite{alba2018entanglementlong}). An important point to stress is that the quasi-particle picture \emph{does not} admit a generalization for describing,
for generic $\alpha$, the growth of R\'enyi entropies~\cite{klobas2021entanglement, bertini2022growth} (with the exception of free systems~\cite{alba2017quench}).

A common starting point for (most of) these results
is the so-called ‘‘replica approach'', whose main idea is that $\text{tr} \rho_A^{\alpha}$ (cf.~\eqref{eq:renyi_def}) can be computed by considering $\alpha$ copies (with $\alpha$ an integer) of the original model, ending up with a ‘‘replicated'' theory.
Appropriate analytic continuation to $\alpha\in\mathbb{R}^{+}$,
gives the R\'enyi and the Von Neumann entropy (see, e.g., \cite{calabrese2004entanglement}).

In particular, within this approach, powerful tools are the so-called
\emph{branch point twist fields}, $T^{\alpha}$ and its hermitian conjugate
$\bar{T}^{\alpha}$. 
Twist fields, in general, are special fields associated to a given symmetry of the theory; they
exist, in a many-body system, for every symmetry transformation. 
The branch point twist fields are special kind of those: as the replicated theory is invariant under permutations of the copies, $T^{\alpha},\bar{T}^{\alpha}$ are the twist fields associated to the generator of cyclic permutations $i\mapsto i+1 \ {\rm mod}\ \alpha$ and its inverse, respectively. The quantities $\text{tr} \rho_A^{\alpha}$ can be related to correlation functions of such twist fields, as first pointed out in quantum field theory in \cite{cardy2008form} clarifying ideas from \cite{calabrese2004entanglement}, and as shown in quantum chains in \cite{castro2011permutation}.

In this work, we make use of the
large-deviation theory for ballistic transport, dubbed ballistic
fluctuations theory (BFT), introduced in \cite{doyon2020fluctuations,myers2020transport}, in order to study the R\'enyi entanglement entropy. The BFT, which is based on hydrodynamic projection principles, gives access to the large-deviation theory for fluctuations of total 2-currents on arbitrary rays in space-time, in homogeneous and stationary states. It generalises in a natural fashion the specific free energy from thermodynamics. By the relation between currents and twist fields, the BFT, as pointed out in \cite{doyon2020fluctuations}, also gives access to two-point functions of twist fields.

Concentrating on (generic) free fermionic systems, we show that both the equilibrium and
the dynamics of R\'enyi entropies at large scales of space and time can be obtained from large-deviation principles and the BFT as applied to branch-point twist fields. The resulting form of the R\'enyi entanglement entropy growth and saturation agree with previous results based on counting particle pairs, but the method is new, and brings out, we believe, important new physics underlying the entanglement entropies. The main two observations are:

(1) We obtain an exact relation between the growth of the R\'enyi entanglement entropies after a so-called integrable \cite{piroli2017integrable,delfino2014quantum,delfino2017theory}, pair-production quench, and static and dynamic ``full counting statistics" in the final GGE. Consider $N_{<,>} := \int_{|v(\theta)|\,<,>\,\xi/2}d\theta\, \psi^\dagger_\theta \psi_\theta$ the conserved quantity giving the total number of ``slow" and ``fast" fermionic modes $\psi_{\theta}$, with speeds $|v(\theta)|<\xi/2$ and $|v(\theta)|>\xi/2$, respectively,
where $\xi=x/t$ is a spacetime ray.
Let us denote $$F_{\rm dyn}^{<,\xi}(\lambda) = \lim_{t\to\infty} t^{-1}\log \braket{e^{\lambda J_{N_<}(t)}}$$ the scaled cumulant generating function for the total current $J_{N_<}(t)$ of slow modes passing through a point in the time interval $[0,t]$ in the final GGE; and $$F_{\rm stat}^{>,\xi}(\lambda) = \lim_{x\to\infty} x^{-1}\log \braket{e^{\lambda N_>(x)}}$$ the scaled cumulant generating function for the total number $N_>(x)$ of fast modes lying on the spatial interval $[0,x]$ in the final GGE. Consider for simplicity $\alpha$ to be even. Then, as $x,t\to\infty$ with $x/t=\xi$ fixed, the R\'enyi entanglement entropy on the interval $[0,x]$, at time $t$ after the quench, has asymptotic form:
\begin{equation}\label{eq:mainfluctu}
    S_\alpha(x,t) \sim
    \frac1{1-\alpha}\Big[
    2 t \sum_{q=-\alpha/2+1}^{\alpha/2} F_{\rm dyn}^{<,\xi}(\ii h_{2q-1})
    \ +\  x \sum_{q=-\alpha/2+1}^{\alpha/2} F_{\rm stat}^{>,\xi}(\ii h_{2q-1})
    \Big],\quad h_p = \frac{\pi p}{\alpha}\ .
\end{equation}
This extends earlier observations of the connection between entanglement entropy and full counting statistics \cite{PhysRevLett.102.100502,PhysRevB.85.035409,Calabrese_2012} to non-equilibrium quenches. Our calculations also provide a fundamental explanation of such relations in terms of twist fields and the large-deviation theory for their asymptotic behaviours, which, as far as we know, has not been noticed before.

(2) We give a new exact derivation of the so-called quasi-particle picture in the case of free fermions with generic dispersion relation. Our derivation is completely independent from the other exact result for the Ising model in \cite{fagotti2008evolution}, which was based instead on Toeplitz matrix representation and multidimensional phase methods. In particular, our method makes transparent how it is the simple structure of long-range correlations induced by particle pairs in integrable quenches that allows one to describe both the growth and saturation of entanglement in a simple and universal way in terms of the long-time GGE, as this structure allows the separation of the contributions of fast and slow modes as per \eqref{eq:mainfluctu}. The emphasis on the structure of long-range correlations also gives a clear understanding as to why for quenches starting from more complicated states, for instance producing correlated groups of more than two particles, more information about the initial state is needed to describe the entanglement growth; in these case no simple formula exists (as showed for example in \cite{bertini2018entanglement,Bastianello_2018}).

We concentrate on free fermion models for simplicity and in order to most clearly illustrate the method and physics. However, as the method is based on general large-deviation and hydrodynamic principles, it is expected to be much more widely applicable, which we leave for future works. In particular, it suggests that hydrodynamic modes and hydrodynamic projections are the more accurate notions at the root of the large-scale behaviour of the entanglement dynamics, rather than particles and their productions.

The paper is organized as follows. Sec.~\ref{sec:BFT_Twist} is an introduction to BFT and its relation to twist fields. In Sec.~\ref{sec:entanglement_twist} we review the replica approach and the associated branch-point twist fields, and we discuss the simplifications occurring in the free fermionic case. Sec.~\ref{sec:entropies_BFT} is the core of the paper, where we derive an expression for correlation function of twist fields from BFT, both in- and out-of- equilibrium, and use them to obtain \eqref{eq:mainfluctu} and recover the known formulas for R\'enyi and entanglement entropies. A discussion of our method and results is given in Sec.\ref{sec:discussion}.
The appendices complement the main text with observations and details of the calculations. In particular, App.~\ref{app:remarks} contains remarks on notions of locality and twist fields. App.~\ref{app:correlations} contains all the details about the applicability of BFT in the different situations we consider, by explicitly computing correlations, and their long-range behaviour, after a quench from a state with pair structure. Finally, App.~\ref{app:Salpha_matrix} is about the structure of the S-matrix in the $\alpha$-copy theory.

\section{Ballistic Fluctuation Theory and twist fields} \label{sec:BFT_Twist}

The BFT \cite{doyon2020fluctuations, myers2020transport}, detailed below, is a theory describing the large-scale, ballistic fluctuations. It is expected to apply to a large class of quantum and classical many-body, extensive systems. It applies to generic systems with space-translation invariant dynamics and interaction range that is short enough. It has been developed originally for states that are spacetime stationary and clustering in space, but many of the ideas have been extended to more general situations \cite{perfetto2021euler,doyon2022ballistic}.

In this paper, we use the BFT as originally developed in \cite{doyon2020fluctuations, myers2020transport},and show how, and under which assumptions, it can be applied to states that emerge after quantum quenches as well. Quantum quenches give rise to states that are locally spacetime stationary, but present time-varying long-range correlations, as we will explain below. We will explain how simple ideas based on the principles explained in \cite{doyon2020fluctuations} allow us to nevertheless use the BFT. We mention that long-range correlations can also, in principle, be accounted for directly by using the more sophisticated ballistic macroscopic fluctuation theory (BMFT) \cite{doyon2022ballistic}, which we leave for further studies.

\subsection{General setting\label{subsec:setting}}

The main strength of the BFT is that it stipulates that only some emergent properties of the system are required in order to describe the large-scale, ballistic fluctuations: the data of its Euler hydrodynamics. We assume the system of interest to
have a certain number $N$ (which in our application to free fermions will be infinite, as the system is integrable) of conserved quantities
\begin{equation}
    Q_{i}=\int dx\, q_{i}(x,t)    
\end{equation}
such that $dQ_{i}/dt=0$. They are assumed to be hermitian (in quantum systems), or real (in classical systems). They include the Hamiltonian $H = \int dx\,h(x,t)$, which generates time translations. For simplicity of the discussion we assume these to be in involution, $[Q_{i,}Q_{j}]=0$ for all
$i,j\in\left\{ 1,\cdots,N\right\} $, however this is not necessary in general. They have associated conservation
laws $\partial_{t}q_{i}+\partial_{x}j_{i}=0$. The observables $q_{i},j_{i}$
are the corresponding charge density and current, assumed to be ``local". In the present paper, locality of $q_i$ and $j_i$ simply means that $Q_i$ has appropriate extensivity properties; we keep the general discussion formal in order to avoid technicalities, but see the remark about locality concepts in the literature in App.~\ref{app:remarks}.

Within such systems, we focus on states belonging to
the manifold of maximal entropy states (MES). Each state is characterized
in terms of a vector $\underline{\beta}=\left\{ \beta_{1},\cdots,\beta_{N}\right\} $ of ``Lagrange multipliers",
with $N$ components (there are as many number of components as the number of
conserved quantities $Q_{i}$). Given such a vector, the density matrix
defining the system reads\footnote{In \eqref{eq:GGE}, an infinite-volume limit needs to be taken. In quantum spin chains, it is shown that if the weight determining the density matrix, $\sum_i\beta_iQ_i$, is short-range, then this defines a state that is unique and exponentially clustering in space \cite{bratteli2012operator}. More generally, one can construct states using the Hilbert space of extensive charges, see \cite{doyonpseudo}.}
\begin{equation}\label{eq:GGE}
    \rho_{\text{\ensuremath{\underbar{\ensuremath{\beta}}}}}\propto e^{-\sum_{i}\beta_{i}Q_{i}}.
\end{equation}
These include the GGEs studied in integrable systems. We consider the system to be in infinite volume. Below, when
no ambiguity occurs, expectation values on such states will be denoted simply as $\langle\cdot\rangle$. Importantly, we assume such states to be clustering strongly enough: connected correlations tend to zero quickly enough at large spatial separations, 
\begin{equation}\label{clustering}
    \langle a(x,0)b(y,0)\rangle^{\rm c}
    :=
    \langle a(x,0)b(y,0)\rangle -
    \langle a(x,0)\rangle \langle b(y,0)\rangle \to 0 \quad (|x-y|\to\infty).
\end{equation}
Here and below, $a(x,t)$ is a local observable at the position $x$ and evolved to time $t$.

As per basic principles of statistical mechanics, the averages of densities are generated by the free energy
\begin{equation}
    \langle q_i\rangle = \frac{\partial}{\partial\beta_i}f(\underline \beta)
\end{equation}
and the mapping
\begin{equation}\label{qbeta}
    \langle \underline q\rangle \leftrightarrow \underline \beta
\end{equation}
is bijective (in appropriate regions of values of $\langle q_i\rangle$ and $\beta_i$).


As mentioned, states that are spacetime stationary for local observables, but with spacetime varying long-range correlations, arise naturally in quantum quenches even after long times. This is because for thermalisation to happen on large regions of the system takes a long time. In such cases, the state is not described by the MES \eqref{eq:GGE}. A precise description of states with long-range correlations is more involved, see e.g.~\cite{doyon2022ballistic,doyon2022emergence}. But the concept of MES is still useful in these situations, as, nevertheless,  averages of all local observables, or observables supported on regions {\em smaller than the correlation range}, still are described by \eqref{eq:GGE}. We will explain how to use this fact in order to ``avoid'' long-range correlations and apply the results of the BFT.


\subsection{Large deviation theory of currents}


Consider some conserved quantity $Q=Q_{i^{*}}$ (for a given $i^{*}\in\left\{ 1,\cdots,N\right\} $),
with associated density $q=q_{i^{*}}$ and current $j=j_{i^{*}}$.

It is instructive to start with a description of the large-deviation theory for extensive charges at equilibrium, before discussing currents. In a given state $\rho_{\underline{\beta}}$, a natural question is to characterize the restriction $\Delta J(\underline x)=-\int_{0}^{x}dx'\:q(x',0)$ of the charge $Q$ to a spatial interval $[0,x]$, and its fluctuations within this interval. Here $\underline{x}$
denotes the horizontal path from $(0,0)$ to $(x,0)$ (and the notation $\Delta J(\underline x)$ is adapted to generalising to currents, as done below). The fluctuations are fully
characterized by the cumulants of $\Delta J(\underline{x})$. It is a simple result that the cumulant generating function at large $x$ is given by a difference of specific free energies $f(\underline\beta)$ of the system,
\begin{equation} \label{deltaf}
    \langle e^{\lambda \Delta J(\underline{x})}\rangle \asymp
    e^{-x \Delta f(\lambda)},\quad \Delta f(\lambda) = f(\{\beta_i + \delta_{ii_*}\lambda\}_i)
    - f(\underline \beta).
\end{equation}
Here and below, we use the notation $A(s)\asymp B(s)$ with the meaning that
$\lim_{s\to\infty}\frac{\log A(s)}{\log B(s)}=1$. Therefore
\begin{equation}
    \Delta f(\lambda) = -\lim_{x\to\infty}
    x^{-1} \log \langle e^{\lambda \Delta J(\underline{x})}\rangle,
\end{equation}
and this generates the scaled cumulants $c_{m}$,
\begin{equation}
    \Delta f(\lambda) =\sum_{m=1}^{\infty}\frac{\lambda^{m}}{m!}c_{m},\quad
    c_{m} = -\lim_{x\to\infty} x^{-1} \langle \, \big[\Delta J(\underline x)\big]^m\,\rangle^{\rm c}.
\end{equation}

When studying transport, similarly, we are interested
in characterizing the total current passing by a given spatial point (e.g.,
the origin) in a given interval of time $[0,t]$. One is therefore interested in the total transfer of $Q$ in time $t$, i.e., $\Delta J(\underline{t})=\int_{0}^{t}dt'\:j(0,t')$,
where now $\underline{t}$ denotes the vertical path from $(0,0)$
to $(0,t)$. As argued in \cite{doyon2020fluctuations} using hydrodynamic principles, the structure parallels closely the equilibrium case in ballistic systems, such as those admitting many conserved charges. At large times $t\to\infty$, a large-deviation principle holds generically for linear scaling with $t$. 

In fact, one can go further and consider the 2-current $\underline{j}=(j,q)$, and the
integral along a more general path $\underline{\ell}$, starting
in $(0,0)$ and ending in $(x,t)$, over its perpendicular component
to the path. This defines the following object
\begin{equation}
\Delta J(\underline{\ell})=\int_{(0,0)}^{(x,t)}\underline{j}(x',t')\land\underline{d\ell}\label{eq:Delta_j}
\end{equation}
where $\underline{d\ell}=(dx',dt')$ and $\underline{j}\land\underline{d\ell} = jd t'- q d x'$. By current
conservation, the integral (\ref{eq:Delta_j}) is in fact independent of the
path chosen, and therefore the result only depends on the end-points $(0,0)$ and $(x,t)$ (for lightness of notation, we keep implicit the dependence on $(0,0)$)
\begin{equation}
    \Delta J(\underline{\ell})=\Delta J(x,t).
\end{equation}
For example, we may choose to connect the initial and final points of
the path via a segment of ray 
\begin{equation}
    \frac xt=\tan\gamma, \quad |\gamma|<\frac \pi2,
\end{equation}
as will be done below. 

Let us consider the Euclidean distance between the initial and final points, 
\begin{equation}
    \ell=\sqrt{x^{2}+t^{2}}
\end{equation}
(we do not assume Euclidean spacetime symmetry, this is simply a convenient way of controlling the scale of $x$ and $t$). As mentioned, in ballistic systems a large-deviation principle holds generically for $\Delta J(x,t)\propto\ell$.
%
Then, as in the equilibrium case, we can define the \emph{scaled
cumulant generating function} (SCGF) $F(\lambda)$ of $\Delta J(x,t)$ as
\begin{equation}
\langle e^{\lambda\:\Delta J(x,t)}\rangle
\asymp e^{\ell F(\lambda)},\quad
F(\lambda)=\lim_{\ell\to\infty}\ell^{-1}\log\langle e^{\lambda\:\Delta J(x,t)}\rangle=\sum_{m=1}^{\infty}\frac{\lambda^{m}}{m!}c_{m}(\gamma) \label{eq:F_lambda}\ .
\end{equation}
The function $F(\lambda)$ also depends on the angle $\gamma$, which we keep implicit.
The coefficients $c_{m}(\gamma)$ are the scaled cumulants of $\Delta J(x,t)$, 
\begin{equation}
    c_m(\gamma) = \lim_{\ell\to\infty} \ell^{-1}
    \langle \big[\Delta J(x,t)\big]^m\rangle^{\rm c},
\end{equation}
which are $m$-point correlation functions of currents and densities integrated over the path $\underline \ell$. The finiteness of scaled cumulants depends on the asymptotic behavior of density and current correlation functions at large spacetime separations. See App.~\ref{app:correlations} for more details.


\subsection{$F(\lambda)$ in MES: biased measure, flow equation} \label{subsec:Flambda}

From the explanations above, at $\gamma = \pi/2$ (in the spatial direction) we have that $F(\lambda) = -\Delta f(\lambda)$, explicitly given in terms of thermodynamic quantities (cf. \eqref{deltaf}). What is the corresponding quantity for finite values of $\gamma$ (i.e., involving the time direction)? It turns out that the answer is completely given using the data of the Euler hydrodynamics of the model.

The Euler hydrodynamics controls the motion and correlations of the many-body system at large scales of space and time. For our purposes, it is sufficient to recall that it is completely fixed by the \emph{flux
jacobian matrix}, defined as (using the bijection \eqref{qbeta})
\begin{equation}
    A_{ij}=\frac{\partial\langle j_{i}\rangle}{\partial\langle q_{j}\rangle}.
\end{equation}
In integrable systems, the flux Jacobian is in fact an infinite dimensional matrix, or more precisely an integral operator~\cite{doyon2020lecture}.

From the definition
it is clear that $A_{ij}$ is basis-dependent. Its fundamental
information is contained in its spectrum $\left\{ v_{k}^{{\rm eff}}\right\} _{k=1,\cdots,N}$, which is composed of eigenvalues if $N$ is finite, and which admits a continuum in integrable systems (where $N=\infty$). The spectrum is interpreted as the set of effective velocities, or ``generalized sound velocities,'' associated to normal modes of the hydrodynamics \cite{doyon2020lecture}.

In order to compute $F(\lambda)$ in (\ref{eq:F_lambda}), the idea
is to bias the measure $\rho_{\underline{\beta}}$ defining the 
MES (where expectation values are considered), by multiplying it by
the exponential of the integrated 2-current $\Delta J(x,t)$, as
appears in (\ref{eq:F_lambda}). The state thus becomes $\lambda-$dependent, and it is in fact a MES, which we write as $\rho_{\lambda;\text{\ensuremath{\underbar{\ensuremath{\beta}}}}}$.
Associated to the segment of path from $(0,0)$ to $(x,t)$, with $\tan\gamma=x/t$,
one can derive a flow equation for $\rho_{\lambda;\text{\ensuremath{\underbar{\ensuremath{\beta}}}}}$, within the space of MES, starting from $\rho_{0;\text{\ensuremath{\underbar{\ensuremath{\beta}}}}}=\rho_{\underline{\beta}}$. Conveniently, this can be written as a flow equation for the Lagrange multipliers
$\underline{\beta}(\lambda;\gamma)$ themselves as follows~\cite{doyon2020fluctuations}
\begin{equation}
\frac{\partial}{\partial\lambda}\beta_{j}(\lambda;\gamma)=-{\rm sgn}\left[A(\lambda;\gamma)-\tan\gamma\:\mathbb{I}_{N}\right]_{i^{*}j},\quad
\beta_j(0;\gamma) = \beta_j
\label{eq:flow}
\end{equation}
where we explicitly introduced the dependence on $\lambda$ as well
as on the path (through $\gamma$) in all quantities.

The main result of BFT is that, solving the flow (\ref{eq:flow}),
one can get an expression of the SCGF directly in terms of the 2-current
$\underline{j}$ evaluated along the flow, namely
\begin{equation}
F(\lambda)=\int_{0}^{\lambda}d\lambda'\left(\cos\gamma\langle j\rangle_{\lambda'}-\sin\gamma\langle q\rangle_{\lambda'}\right)\label{eq:F_lambda-3}
\end{equation}
where $\langle\cdot\rangle_{\lambda}$ denotes expectation values on
$\rho_{\lambda;\underline{\beta}}$ (recall that $\langle\cdot\rangle_{0}\equiv\langle\cdot\rangle$) and $\langle j\rangle_{\lambda'} = \langle j(0,0)\rangle_{\lambda'}$, $\langle q\rangle_{\lambda'}=\langle q(0,0)\rangle_{\lambda'}$.
Crucially,
from Eq. (\ref{eq:F_lambda-3}), one has that $F(\lambda)$ is given
in terms of thermodynamic and Euler hydrodynamic objects only. 

The result \eqref{eq:F_lambda-3} with \eqref{eq:flow} follows from a large-deviation principle. The full
derivation of Eqs. (\ref{eq:flow}-\ref{eq:F_lambda-3}) is not reported
here, but can be found in \cite{doyon2020fluctuations}.

The main assumption underlying the validity of \eqref{eq:F_lambda-3}
is that of \emph{``strong enough'' clustering along
the space-time ray of velocity} $x/t=\tan\gamma$. Specifically, spatial clustering \eqref{clustering} is not enough: one needs vanishing of correlation functions of perpendicular currents $j_\perp(x,t) := \underline j(x,t)\wedge \underline{d\ell}$, the integrand in \eqref{eq:Delta_j}, when the distance along the ray $(0,0)\to(x,t)$ goes to infinity,
\begin{equation}\label{clustering2}
    \langle j_\perp(y,s)j_\perp(y+\ell\sin\gamma,s+\ell\cos\gamma)\rangle^{\rm c}
    \to 0 \quad (\ell\to\infty, y/s = \tan \gamma)
\end{equation}
and similar requirements for all multipoint functions of perpendicular currents. The vanishing must be fast
enough to make integrals defining the cumulants rapidly converging.

A crucial remark for our work, concerning the requirement of clustering, is Remark 3.3 of \cite{doyon2020fluctuations}. Recall that $\Delta J(x,t) = \Delta J(\underline\ell)$ in \eqref{eq:Delta_j} is independent of the path $\underline\ell$, and only depends on the end-points $(0,0)$ (which we have kept implicit) and $(x,t)$. One may therefore hope to apply the general result of the BFT for each path element and obtain, in place of \eqref{eq:F_lambda-3}, the expression
\begin{equation}
    F(\lambda) = \lim_{\ell\to\infty} \ell^{-1}\int_0^\lambda
    d\lambda'\,\int_{(0,0)}^{(x,t)}
    \langle\underline j(x',t')\rangle_{\lambda'}\wedge\underline{d\ell},\label{eq:F_lambda-4}
\end{equation}
for any path $\underline\ell$ (with a well-defined large-scale limit $\ell\to\infty$). As explained in \cite[Rem 3.3]{doyon2020fluctuations}, this is expected to be correct {\em if and only if there are no strong correlations between the perpendicular currents amongst different points on the path}. 

Eq. \eqref{eq:F_lambda-4} is the main result from BFT which will be applied below to expectation values of observables (specifically, of twist fields) both at equilibrium in states described by GGEs (in Sec. \ref{subsec:spacedirection}), and out-of-equilibrium in states emerging after quenches (in Sec. \ref{ssect:longrange} and the following ones). In the latter case, this can be done by choosing a path that ``avoids" the long-range correlations that such states present, so that expectation values in the pre-quench state can be approximated with the ones in the corresponding long-time GGE (where correlations can be shown to decay fast enough). 

\medskip \noindent
{\em Remark (clustering and the ballistic large-deviation principle).} If there is no path that satisfies strong clustering, then typically the ballistic large deviation principle is broken, and the SCGF resulting from the BFT may be infinite or zero. Exponential clustering, which is sufficiently strong, is expected to hold on rays away from the fluid velocities, $\tan\gamma \neq v^{\rm eff}_i\,\forall i$, in generic equilibrium states. In GGEs of integrable systems at nonzero entropy density, the spectrum of the flux Jacobian contains a continuum, and clustering is in fact a power-law, with power $1/\ell^2$ for the perpendicular currents, Eq.~\eqref{clustering2}, for all rays within this continuum, see App.~\ref{app:global} for the case of free fermions. This is still strong enough for the BFT results to hold, as confirmed numerically for the hard rods \cite{myers2020transport}. By contrast, in non-integrable systems, where the flux Jacobian has a discrete spectrum, clustering is too weak on rays along any of the eigenvalues of the spectrum of the flux Jacobian (fluid velocties), $\tan\gamma = v^{\rm eff}_i$ for some $i$. In these directions, the ballistic large-deviation principle is broken. See the discussion in \cite{doyon2020fluctuations}. In general, at zero temperatures when there is no gap, weak power-law clustering is seen and the ballistic large-deviation principle is also broken. When the breaking of the large-deviation principle happens as we change a parameter (a (generalised) temperature, a coupling), this can be seen as a ``dynamical phase transition".

\subsection{Application of the BFT
to twist fields} \label{subsec:twistfields}

One of the most important application of the BFT is to two-point correlation
functions of so-called ``twist fields''. This is useful, because, as explained in the introduction, twist fields are probably the most efficient way of studying entanglement entropy, the main object of this work.

There are a number of ways of defining twist fields, and we will discuss two natural approaches. The first is natural in the context of the large-deviation theory as recalled above and based on the explicit knowledge of extensive conserved quantities; it applies to classical and quantum systems alike. The second is a more abstract formulation that does not require the explicit knowledge of extensive conserved quantities, but that is better adapted to quantum systems.

Consider as above an extensive conserved quantity $Q$. Recall that $Q$ has associated density and current $q(x,t)$ and $j(x,t)$. It is convenient to define the {\em height field} $\varphi(x,t)$ via the relations
\begin{equation}
q(x,t)=\partial_{x}\varphi(x,t),\qquad j(x,t)=-\partial_{t}\varphi(x,t).
\end{equation}
This ensures that the continuity equation $\partial_{t}q+\partial_{x}j=0$ is automatically satisfied. The height field is unbounded (because the charge is extensive), and we note in particular that differences grow linearly,
\begin{equation} \label{deltavarphiDeltaJ}
\varphi(0,0)-\varphi(x,t) = \Delta J(x,t) \propto \ell.
\end{equation}
The height field may be written as an integral over half space\footnote{This expression is somewhat formal. In quantum spin chains, by applying a time derivative to \eqref{twistfieldintegral}, one can make the result mathematically rigorous using an appropriate Hilbert space of the Gelfand-Naimark-Segal type, as proved in \cite{doyonlinearised}.},
\begin{equation}\label{twistfieldintegral}
    \varphi(x,t) = -\int_{x}^{\infty} dy\,
    q(y,t) + \varphi(\infty,t).
\end{equation}

One finds that the boundary term at infinity is constant in time, $\varphi(\infty,t) = \varphi(\infty)$ (which can be chosen to vanish). Further, it is clear that the result is independent from the choice of path in spacetime thanks to the conservation laws,
\begin{equation}
    \varphi(x,t) = \int_{(x,t)}^{(\infty,t)} \underline j\wedge \underline{d\ell}
    +\varphi(\infty),
\end{equation}
and thus the height field $\varphi(x,t)$ only depends on the point $(x,t)$. This justifies the notation. One may also choose a different direction for the half-space integral, the difference being encoded within
\begin{equation}
    \varphi(\infty)-\varphi(-\infty)=Q.
\end{equation}

Exponentials of $\varphi(x,t)$, that is 
\begin{equation} \label{Tvarphi}
    T_{\lambda}=e^{\lambda\varphi},
\end{equation}
are known in general as ``twist fields''. Because of the expression \eqref{twistfieldintegral}, they are not ``local" in the na\"ive sense, but are usually referred to as ``semilocal" in the literature (see App.~\ref{app:remarks} for an overview); this is made clearer below using exchange relations.  As an immediate result of the large-deviation analysis, using
\begin{equation} \label{TTexpvarphi}
T_{\lambda}(0,0)T_{-\lambda}(x,t) = e^{\lambda( \varphi(0,0)-\varphi(x,t))}   
\end{equation}
with \eqref{deltavarphiDeltaJ}, we get the leading exponential behavior of the two-point correlations of twist fields as
\begin{equation}
\langle T_{\lambda}(0,0)T_{-\lambda}(x,t)\rangle\asymp e^{\ell F(\lambda)}\ .\label{eq:bft_twist}
\end{equation}
That is, if the ballistic large-derivation principle holds for the charge $Q$, then the associated twist field shows an exponential behaviour at large space-time separations. Because $e^{\lambda\varphi}$ is not bounded for $\lambda\in\mathbb R$, $T_\lambda$ should be referred to as an ``unbounded twist field".

In quantum models, because of the lack of commutativity of observables, and {\em a fortiori} in quantum field theory (QFT), because, additionally, of the necessary renormalisation procedure applied to the twist fields, the relation \eqref{TTexpvarphi} is not strictly valid. However, corrections do not affect the leading exponential behaviour of the two-point function, as argued for the XX quantum chain in \cite{del2022hydrodynamic}.

A second viewpoint on twist fields is as follows. A twist field $T(x,t)$ is in general an operator associated to a symmetry transformation that ``acts locally enough". This is a transformation $a(x,t) \mapsto \tilde a(x,t)$ of the model, that maps local observables to local observables, and that preserves the operator algebra; one usually also requires that it preserves the Hamiltonian density, $\tilde h(x,t)= h(x,t)$. The property of an operator $T(x,t)$ that makes it a twist field associated to such a symmetry, is the following equal-time exchange relation:
\begin{equation}\label{exchangerelationgeneral}
    T(x,t) a(y,t) = \left\{\begin{array}{ll}
         \tilde a(y,t) T(x,t)& y\gg x \\
         a(y,t) T(x,t) & y\ll x
    \end{array}\right.
\end{equation}
for every local observable $a(y,t)$. Corrections at larges distances $|y-x|$ should be small enough, for instance exponentially decaying. This equal-time exchange relation is known in the literature as expressing the ``semilocality" of the twist field\footnote{If the twist field indeed preserves the hamiltonian density, it commutes with it at large distances, up to small (e.g.~exponential) corrections. This has important implications, which justifies considering twist fields, for many purposes, on the same footing as local fields; for instance, the time-evolved twist field is analytic in the time variable at small enough times.} $T(x,t)$.

In general, symmetry transformations act via unitary operators $U$ as $\tilde a(x,t) = U a(x,t) U^{-1}$. For instance, in quantum spin chains, one often can write $U = \prod_{x\in\mathbb Z} U_x$ where $U_x$ acts non-trivially on a neighbourhood of $x$ (possibly up to exponentially decaying corrections); this is indeed local enough. In such instances, one can simple set
\begin{equation}\label{eq:twistU}
    T(x,0) = \prod_{y\geq x}
    U_y.
\end{equation}

How does the exchange-relation formulation \eqref{exchangerelationgeneral} connect with the height-field formulation \eqref{Tvarphi} of twist fields? This is from the general principle that {\em every extensive conserved quantity gives rise to a continuous, one-parameter unitary group of symmetry transformations that act locally enough}. That is, given an extensive $Q$ (recall that it is assumed to be hermitian), we may consider the symmetry transformation
\begin{equation}\label{eq:tildtransfo}
    \tilde a(x,t) = e^{\ii \eta Q} a(x,t) e^{-\ii \eta Q}
\end{equation}
for a real parameter $\eta \in\mathbb R$. It is simple to see that this acts locally enough, as described above. Indeed, by conservation we can replace $Q$ by $Q(t)$ in \eqref{eq:tildtransfo}, and by locality of the density $q(x,t)$, we have that $\int_{-L}^L dx\,[q(x,t),a(0,t)]$ approaches its limit as $L\to\infty$ quickly enough, and gives in the limit a local observable supported at $x=0$ at time $t$. Therefore, by using the Baker-Campbell-Hausdorff formula, at least for all $\eta$ small enough, $e^{\ii \eta Q} a(x,t) e^{-\ii \eta Q}$ gives rise to a local observable at $(x,t)$. Using the same principles, it is then a simple matter to show that \eqref{exchangerelationgeneral} holds with the choice \eqref{twistfieldintegral} for the height field, and the identification in \eqref{Tvarphi}
\begin{equation}\label{eq:identificationtwist}
    T = T_{-\ii\eta}.
\end{equation}
Because $e^{-\ii \eta \varphi}$ is bounded, we will referred to these as ‘‘bounded twist fields''. Bounded twist fields are the ones usually considered in the literature. In particular, because they are bounded, their two-point functions should decay in spacetime,
\begin{equation}
    \Re F(-\ii\eta)\leq 0.
\end{equation}

One may in fact argue that, viceversa, {\em to every symmetry transformation that acts locally enough, we can associate an extensive conserved quantity.} In quantum field theory, Noether's theorem shows that, for continuous symmetry groups, we indeed have $U = e^{\ii \eta Q}$ for some conserved quantity $Q$ associated to a conserved 2-current; again this is local enough. In general, for any local enough transformations, one can identify, formally, an extensive charge $Q$ with the operator $-\ii\log U$ (thus taking \eqref{eq:tildtransfo} with $\eta=1$), a formal construction that, we expect, could be used fruitfully within the BFT. In the present paper, we will consider a twist field associated to a discrete symmetry transformation, but this will be embedded within a continuous symmetry group thanks to the free-fermion structure,  thus the charge $Q$ will be explicit.

Finally, we observe that applying the BFT for the bounded twist fields, using the identification \eqref{eq:identificationtwist}, requires an {\em analytic continuation of $\lambda$ in the BFT formulae, to purely imaginary values $-\ii \eta$}. This is a subtle aspect, as for purely imaginary values of $\lambda$, the modified state by the flow equation is not strictly a MES (because the resulting linear functional on the algebra of observables is not necessarily positive). We believe that if fluid velocities are well separated, as is typically the case in non-integrable systems, then the analytic continuation can be obtained meaningfully by keeping the sign of the eigenvalues constant in the flow equation \eqref{eq:flow} (as the analytic continuation will not ``see" the jumps in eigenvalues), and integrating the flow in the complex $\lambda$-plane. In free fermion models, the analytic continuation can be performed directly on the explicit result for $F(\lambda)$, as done in \cite{del2022hydrodynamic} and in the next section. We will show below that the BFT indeed predicts decay of correlation functions in this case. We leave the discussion for interacting integrable systems to future works.

See App.~\ref{app:remarks} for a brief discussion of notions of locality and twist fields.

\subsection{Explicit expression of $F(\lambda)$ in free fermionic theories } \label{ssect:freefermion}

Up to now, the theory was general, and all equations correctly give the ballistic part of large deviations in general systems with the properties detailed in Secs. \ref{subsec:setting} and \ref{subsec:Flambda}. When considering integrable systems, Eq. (\ref{eq:F_lambda-3})
gets further simplified. In fact, using the theory
of generalized hydrodynamics (GHD)~\cite{bertini2016transport,castro2016emergent}, an explicit expression
of the hydrodynamic quantities $\langle q\rangle_{\lambda},\langle j\rangle_{\lambda}$
along the flow can be worked out.

We now overview the simplified expression for $F(\lambda)$ in the special
case of free fermions in the continuum, arguably the easieast among
1D integrable systems, which is our focus in this paper (the corresponding
results in the case of generic interacting integrable models can be
found in \cite{myers2020transport}).


In order to keep the structure general, we simply assume that a fermionic, complex field $\psi(x,t)$ exists with interactions that are quadratic and short-range. Its Fourier modes are denoted $\psi_\theta$, with anti-commutation relation $\{\psi_\theta^\dagger,\psi_{\theta'}\} = \delta(\theta-\theta')$. Here $\theta$ represents the momentum, which we assume takes values in $\mathbb R$ for simplicity (for quantum chains, this would be a bounded subset instead, but the general ideas are not affected). We also denote the dispersion relation as $E(\theta)$, which we assume is strictly convex and symmetric $E(\theta) = E(-\theta)$. Thus, under canonical normalisation,
\begin{equation}
    \psi(x,t) = \frac1{\sqrt{2\pi}} \int d\theta\,e^{\ii \theta x - \ii E(\theta) t} \psi_\theta.
    \label{eq:fourier_transform}
\end{equation}
As it is integrable, the model possesses an infinite number of conserved
quantities. A ``scattering basis" for these is given by $Q_{\theta}=\psi_{\theta}^{\dagger}\psi_{\theta},\,\theta\in\mathbb R$. Strictly speaking, the $Q_{\theta}$'s are not linearly extensive, but (for generic dispersion relation)
any extensive conserved quantity can be obtained by a suitable ``linear combination", or basis decomposition, $Q_i = \int d\theta\,h_i(\theta)Q_\theta$. Here, $h_{i}(\theta)$
is the one-particle eigenvalue of the extensive charge $Q_{i}$. Examples are the number of fermions $N = \int d\theta\,Q_{\theta}$, the total momentum $P=\int d\theta\:\theta Q_{\theta}$, and the total energy $H=\int d\theta\:E(\theta)Q_{\theta}$.
A typical GGE \eqref{eq:GGE} takes the form
$\rho_{w} := \rho_{\underline{\beta}}\propto e^{-\int d\theta\, w(\theta)Q_{\theta}}$,
where $w(\theta)=\sum_{i}\beta_{i}h_{i}(\theta)$ is the generalised Boltzmann weight in the particle basis .
For example, for a thermal state, we have $\rho_{w}
\propto e^{-\beta (H-\mu N)}=e^{-\int d{\theta}\,\beta(E(\theta)-\mu)Q_{\theta}}$,
so $w(\theta)=\beta(E(\theta)-\mu)$.

In general, the physical meaning of the Lagrange multipliers $\beta_i$ depends on the choice of the set of charges $Q_i$, i.e.~on the choice of the set of one-particle eigenvalues $h_i(\theta)$. But there is no need to choose any particular infinite set of charges $Q_i$, or to write explcitly $w(\theta)$ in a basis decomposition $w(\theta)=\sum_{i}\beta_{i}h_{i}(\theta)$. The function $w(\theta)$ fixes the GGE, and only few basic requirements constrain $w(\theta)$ for $\rho_{w}$ to be a valid GGE (we will ask that it be positive and grow sufficiently fast as $|\theta|\to\infty$). We note that the conserved charge densities take the standard form $\langle q_i\rangle = \int d\theta/(2\pi)\,n(\theta) h_i(\theta)$ in terms of the occupation function
\begin{equation}\label{eq:occupation}
    n(\theta)=\frac{1}{1+e^{w(\theta)}}\ .
\end{equation}
and that, in a system of length $L$ with periodic boundary conditions, we have $\langle Q_\theta\rangle = \frac L{2\pi} n(\theta)$.

In our calculations, we will assume that $n(\theta)$ has an analytic extension in a neighbourhood of $\mathbb R$, and that $n(\theta)\to0$ as $|\theta|\to\infty$.

Consider the large-deviation problem for the charge $Q = Q_{i^*}$, with one-particle eigenvalue $h_{i^*}(\theta)= h(\theta)$. For free fermions, $F(\lambda)$ simplifies to
\begin{equation}
F(\lambda)=-\int\frac{d\theta}{2\pi}\left|v(\theta)\:\cos\gamma-\sin\gamma\right|\left[f(\epsilon_{\lambda}(\theta;\gamma))-f(w(\theta))\right]\label{eq:largedev}
\end{equation}
where $v(\theta) = dE(\theta)/d\theta$ is the group velocity\footnote{The effective
velocity of GHD is just the group velocity in free particle models, $v^{{\rm eff}}(\theta)=v(\theta)$.}. The function
$f(\epsilon)$ is the fermionic free energy function (the free energy density per distance and per unit rapidity $\theta$),
\begin{equation}
f(\epsilon)=-\log(1+e^{-\epsilon}),
\end{equation}
and the function $\epsilon_{\lambda}(\theta;\gamma)$ is the Boltzmann weight along the flow \eqref{eq:flow} in the particle basis.
Its initial condition is $\epsilon_{0}(\theta;\gamma)=w(\theta)$, and the corresponding flow equation (which simply follows from \eqref{eq:flow} in terms of $\beta_i$) simplifies to
\begin{equation} \label{flow_epsilon_lambda}
\partial_{\lambda}\epsilon_{\lambda}(\theta;\gamma)=\sgn(\tan\gamma - v(\theta))\,h(\theta)\ ,
\end{equation}
which is explicitly solved as
\begin{equation}
\epsilon_{\lambda}(\theta;\gamma)=w(\theta) + \lambda\,\sgn(\tan\gamma - v(\theta))\, h(\theta)\ .
\end{equation}

As mentioned above, in order to apply the BFT to bounded twist fields associated to symmetry transformations, one needs to perform an analytic continuation in $\lambda$ to the purely imaginary direction $\lambda = -\ii \eta,\,\eta\in\mathbb R$. In free Fermion systems this is simple to do, as the above formulae can be directly analytically continued. The resulting $F(\lambda)$ possesses, in general, both a real and an imaginary part. The real part describes the exponential decay of the two-point correlation functions of twist fields, while the imaginary part describes oscillations. In the following, we will not discuss oscillations, as their full description would require a more in-depth analysis; we will concentrate on the exponential decay, hence the real part of $F(\lambda)$.

Correlation functions of twist fields are expected to be decaying at large spacetime distances. It is simple to show from \eqref{eq:largedev} that indeed\footnote{This is because in \eqref{eq:largedev} one has $e^{-\epsilon_{-\ii \eta}(\theta;\gamma)} = ru$ where $r=e^{-w(\theta)}>0$ and $u$ is a pure phase, $|u|=1$, and $|1+r|\geq|1+ru|$ for any $r>0$ and any pure phase $u$.}, $\Re F(-\ii\eta)\leq 0$.

One important remark is that the only information required about the current $\Delta J(x,t)$ whose SCGF is taken, is {\em the one-particle eigenvalue $h(\theta)$ of the corresponding total charge $Q$}. Thus, the BFT predicts that only a limited amount of information about the twist field is required 
in order to evaluate the exponential asymptotic of its two-point function. Note that this is true also in the interacting case.

\section{Entanglement and branch-point twist fields } \label{sec:entanglement_twist}

In this section, we recall how entanglement entropies can be computed using a certain type of twist fields, called branch-point twist fields, associated to permutation symmetries. We then recall that, in free fermionic theories, these can be re-written in terms of $U(1)$ twist fields. This will be used in the next section in order to apply the BFT to the calculation of entanglement entropies.

\subsection{Replicas and branch-point twist fields}

Within the replica method, in order to compute entanglement entropies (cf. Eqs.~\eqref{eq:renyi_def}-\eqref{eq:entent_def}) in a given theory, one re-writes the quantity ${\rm tr}\rho_A^{\alpha}$ in terms of an appropriate expectation value in the {\em replica model}. This is the model composed of $\alpha$ independent, commuting copies of the original model ($\alpha\in\mathbb{N}$). For a one-dimensional system in a state with density matrix $\rho$, and with the subsystem $A$ being a single interval, e.g., $A=[x_1,x_2]$, it is a simple matter to show \cite{cardy2008form,castro2011permutation} that ${\rm tr}\rho_A^{\alpha}$ is exactly identified with the two-point function of {\em branch-point twist fields},
\begin{equation}
\text{tr}\rho_A^{\alpha} = \langle T^{\alpha}(x_1,0)\bar{T}^{\alpha}(x_2,0)\rangle_{\rho^{\otimes \alpha}}.\label{eq:traces2}
\end{equation}
The expectation value on the r.h.s.~is computed in the density matrix $\rho^{\otimes \alpha} = \otimes_{i=1}^{\alpha} \rho_i$, where $\rho_i$ is the original density matrix, on copy $i$. Branch-point twist fields in the replica theory are twist fields associated to the symmetry under replica cyclic permutations of order $\alpha$ (which generate the group $Z_{\alpha}$). They take the product form \eqref{eq:twistU}, involving on-site copy-permutation operators\footnote{Here we omit any regularisation issue that may arise in models on a continuous space, which, as mention, do not affect exponential asymptotic behaviours.} \cite{castro2011permutation}:
\begin{equation}
    T^\alpha(x,0) = \prod_{y\geq x} P_y
\end{equation}
and $\bar T^\alpha(x,0)= \big(T^\alpha(x,0)\big)^\dagger$. Here, denoting by $a_i(x)$ observables lying on (that is, acting nontrivially only on) copy $i\in\{1,2,\ldots,\alpha\}$ and position $x$, and identifying $a_{\alpha+1}(x)\equiv a_1(x)$, the permutation unitary is defined by
\begin{equation}
    P_x a_i(y)P_x^{-1}
    = \begin{cases} a_{i+1}(y)
    & y=x \\
    a_i(y) & y\neq x.
    \end{cases}
\end{equation}
This implies the equal-time exchange relations (see \eqref{exchangerelationgeneral}) 
\begin{equation}\label{eq:permexch1}
    T^{\alpha} (x,t) a_i (y,t)  = 
    \begin{cases}
    a_{i+1} (y,t) T^{\alpha} (x,t) & y \geq x \\
    T^{\alpha} (x,t) a_{i} (y,t) & y<x
    \end{cases}
\end{equation}
and
\begin{equation}\label{eq:permexch2}
    \bar T^{\alpha} (x,t) a_i (y,t)  = 
    \begin{cases}
    a_{i-1} (y,t) \bar T^{\alpha} (x,t) & y \geq x \\
    \bar T^{\alpha} (x,t) a_{i} (y,t) & y<x \ .
    \end{cases}
\end{equation}
From Eq.~\eqref{eq:traces2}, R\'enyi entanglement entropies
can be simply obtained via Eqs.~(\ref{eq:renyi_def})-(\ref{eq:entent_def}).


In the context of (1+1)-dimensional QFT, exchange relations of the form \eqref{eq:permexch1}, \eqref{eq:permexch2} give the most appropriate formulation for working definitions of the branch-point twist field. It is in this context that they were first introduced \cite{cardy2008form}, as a way of evaluating partition functions on branched surfaces, taking inspiration from \cite{calabrese2004entanglement}.

We note that the action of branch-point twist fields can be diagonalized by going to the Fourier basis in the replica index  (we choose anti-periodic boundary conditions in replica space, see below),
\begin{equation}\label{eq:fouriertransform}
    a_p(x,t) = \mathcal{F}_{i\to p}[a_i(x,t)] := \frac1{\sqrt\alpha}
    \sum_{i =1}^{\alpha} e^{\ii \pi  p i/\alpha} a_i (x,t)
\end{equation}
for $p \in \left\{0,1,\ldots,\alpha-1\right\}$. This gives
\begin{equation} \label{Tpa}
    T^{\alpha} (x,t) a_p (y,t)  = 
    \begin{cases}
    e^{-\ii \pi p/\alpha} a_{p} (y,t) T^{\alpha} (x,t) & y \geq x \\
    a_{p} (y,t) T^{\alpha} (x,t) & y < x
    \end{cases}
\end{equation}
and similarly for $\bar T^\alpha(x,t)$. In the next subsection we will use a similar construction, albeit in a different basis of the replica model.

%
%

%
%
%

As will be explained in Sec.~\ref{sec:entropies_BFT}, for our purposes, the most general object we need to consider is the two-point correlation functions
\begin{equation}
\langle T^{\alpha}(x_1,t_1)\bar{T}^{\alpha}(x_2,t_2)\rangle_{\rho^{\otimes \alpha}}
\end{equation}
at different spacetime points.

\subsection{Enhanced symmetry in free fermions: from $Z_{\alpha}$ to $U(\alpha)$} \label{ssect:enhanced}

Consider the special case of free fermions, see Sec.~\ref{ssect:freefermion}. In this case, because of the quadratic nature of free fermion Hamiltonians, the $Z_{\alpha}$
symmetry of the replicated theory turns out to be embedded into the larger symmetry group $U(\alpha)$, which accounts for not only permutation of replicas, but also rotations amongst them. Thus, the branch-point twist field is a twist field associated to a particular symmetry transformation, part of a continuous symmetry group. As explained in Sec.~\ref{subsec:twistfields}, using Noether's theorem, this then allows one to write an explicit extensive charge associated to the twist field \cite{cardy2008form}.

The $U(\alpha)$ symmetry is most clearly expressed in a {\em different basis} of the replica theory than that used above, obtained by ``fermionising" the replica theory. By the basic construction of the replica theory, different replicas commute with each other. However, in order to extract the symmetry $U(\alpha)$, one needs fermions in different copies to {\em anti-commute}. One simply defines the replica theory by asserting that fermion fields anti-commute. This is of course natural, but changes the action of the branch-point twist field (the exchange relations \eqref{eq:permexch1}, \eqref{eq:permexch2}) by introducing an extra minus sign, as worked out in \cite{cardy2008form}. From now on, we denote
\begin{equation}
    \psi_i(x,t) = \frac1{\sqrt{2\pi}} \int d\theta\,e^{\ii \theta x - \ii E(\theta) t} \psi_{\theta,i}
    \label{eq:fourier_transform_replica}
\end{equation}
the Dirac fermion on the $i$-the copy {\em in this new basis}; thus $\psi_i(x,t)\psi_j(x',t') = -\psi_j(x',t')\psi_i(x,t)$ if $i\neq j$, and the canonical anti-commutation relations hold, $\{\psi_i(x,t),\psi_j^\dag(x',t)\} = \delta_{ij}\delta(x-x')$.

We now recall the main arguments of \cite{cardy2008form} in order to obtain a useful form of the branch-point twist field. In the new basis, the $U(\alpha)$ symmetry is explicitly a linear action on the fermions, in its fundamental representation. Most importantly, in this basis, the cyclic permutation $\psi_i\to \psi_{i+1}$ is  a particular element of $U(\alpha)$, which is in fact an element of a $U(1)$ subgroup. The Fourier transform \eqref{eq:fouriertransform} can be performed in this basis,
\begin{equation}\label{eq:fouriertransformpsi}
    \psi_p(x,t) = \frac1{\sqrt\alpha}
    \sum_{i =1}^{\alpha} e^{\ii \pi  p i/\alpha} \psi_i (x,t).
\end{equation}
This diagonalises that $U(1)$ subgroup; the action of the twist field is then diagonalised. In fact, it turns out that the anti-commuting basis is also the one that guarantees that the Fourier transform operation keeps the $S$-matrix diagonal, see Appendix~\ref{app:Salpha_matrix}. Thus both the charge associated to the twist field, and the $S$-matrix, are diagonal in terms of the particles corresponding to $\psi_p$ -- this is at the root of the simplification.

The fermion fields $\psi_p(x,t)$ after Fourier transform are still independent free fermions with canonical anti-commutation relations. Each Fourier sector admits an independent $U(1)$ symmetry, and, as shown in \cite{cardy2008form}, the branch-point twist field can be written as a product of $U(1)$ twist fields acting nontrivially on each Fourier sector. Because of the extra minus sign in the twist field action, it is simpler to concentrate on the case of $\alpha$ even (the full dependence on $\alpha$ is obtained by analytic continuation), in which case the product goes over the following values of momenta:
\begin{equation}
    p\in I_{\alpha}:=\left\{ -\alpha+1, -\alpha + 3, \cdots,\alpha-1\right\}.
\end{equation}
Specifically, it is found that \cite{cardy2008form}
\begin{equation}
T^{\alpha}=\prod_{p\in I_{\alpha}}\tau_{p}^{\alpha}=\prod_{q=-\alpha/2+1}^{\alpha/2}\tau_{2q-1}^{\alpha}
\label{eq:decomposition}
\end{equation}
with $\tau_{p}^{\alpha}$ being a $U(1)$ twist field acting non-trivially
only on $\psi_{p}$ (as a phase),
\begin{equation} \label{Tp}
    \tau_p^{\alpha} (x,t) \psi_q (y,t)  = 
    \begin{cases}
    e^{-\ii \pi p/\alpha} \psi_{q} (y,t) \tau_p^{\alpha} (x,t) & y \geq x \mbox{ and } p=q \\
    \psi_{q} (y,t) \tau_p^{\alpha} (x,t) & y < x \mbox{ or } p\neq q
    \end{cases}
\end{equation}
(cf.~\eqref{Tpa}).

The decomposition \eqref{eq:decomposition} allows us to factorise the branch-point twist field two-point functions into products of $U(1)$ twist field two-point
functions. This however only holds if the state can be likewise factorised. This is nontrivial: the state $\rho^{\otimes \alpha}$ is naturally factorised in copy space, but not necessarily in the Fourier-copy space. It is a simple matter to verify that if $\rho$ satisfies Wick theorem, then $\rho^{\otimes \alpha}$ also factorises as a tensor product of states $\rho$ in Fourier-copy space; this is because such states are completely determined by fermion two-point functions, which stay diagonal in Fourier-copy space. Therefore, we have, in Wick-theorem states $\rho$,
\begin{equation}
\langle T^{\alpha}(0,0)\bar{T}^{\alpha}(x,t)\rangle_{\rho^{\otimes \alpha}}=\prod_{q=-\alpha/2+1}^{\alpha/2}\langle\tau_{2q-1}^{\alpha}(0,0)\bar{\tau}_{2q-1}^{\alpha}(x,t)\rangle_\rho
\ .\label{eq:U1_twist}
\end{equation}
Note how on the right-hand side, each factor is evaluated in the state $\rho$ for the fermion $\psi_{2q-1}$.

In the following we are going to apply the BFT machinery to each correlation function of the $U(1)$ twist fields.
The crucial fact that makes it simple is that, for any given $p$, $\tau_{p}^{\alpha}(x,t)$ is the (bounded) twist field associated to the $U(1)$-charge
\begin{equation}
Q_{p} = \frac{\pi p}\alpha \int  dx\,\psi^\dagger_p(x)\psi_p(x)
= \frac{\pi p}\alpha \int d\theta\,
\psi_{\theta,p}^\dag \psi_{\theta,p}\ ,
\end{equation}
with explicit expressions as exponential of half-space integrals of charge densities, as per Eq.~\eqref{Tvarphi}:
\begin{equation}\label{eq:tauexp}
    \tau_p^\alpha(x,t)
    = \exp\Big[
    \ii \int_{x}^{\infty} dx'\,
    q_p(x',t)
    \Big],\qquad q_p(x,t) = \frac{\pi p}\alpha  \psi_p^\dag(x,t)\psi_p(x,t)\ .
\end{equation}
$Q_p$ acts on the single-particle basis as
\begin{equation}
Q_{p}|\theta,q\rangle=h_{p}\delta_{p,q}|\theta,q\rangle, \quad {\rm with} \; h_{p}= \frac{\pi p}\alpha 
\label{eq:Qalphap}
\end{equation}
(note that $\psi_p(x)$ has $Q_p$-charge $-h_p$, in agreement with \eqref{Tp}). With $Q=Q_p$, the twist field $\tau_p^\alpha$ is identified with $\tau_p^\alpha = T_{-\ii }$ in the notation of \eqref{eq:identificationtwist} (that is, with $\eta=1$), acting on the sector $p$. Recall that the action of the charge on the single-particle basis is all we need to know in order to apply the BFT (cf. \eqref{eq:largedev} and \eqref{flow_epsilon_lambda}).

\section{Entanglement entropies from BFT}
\label{sec:entropies_BFT}

We arrived to a rewriting of the two-point function of the branch-point
twist fields as product of two-point functions of $U(1)$ twist fields,
Eq.~(\ref{eq:U1_twist}). From there, using BFT, all such components
can be accessed via Eq.~(\ref{eq:bft_twist}) specified to the twist
fields $\tau_{p}^{\alpha}$, which in the notation of Eq.~(\ref{eq:bft_twist}) is identified with $T_{-\ii }$, with the r.h.s.~evaluated via Eq.~(\ref{eq:largedev}),
thus exploiting the free fermionic nature of the problem. Considering
different choices of the points in spacetime where the global fields
$T^{\alpha},\bar{T}^{\alpha}$ are located, we are able to access
R\'enyi entropies both at equilibrium and after a quench, as we are
now going to discuss.

\subsection{R\'enyi entropies of a finite interval in a GGE (charge fluctuations in space)}
\label{subsec:spacedirection}

We start by considering the $\alpha-$R\'enyi entropy of a finite interval
$A=[0,x]$ within a generic GGE $\rho_w$ uniquely defined by the function
$w(\theta)$ (see Sec.~\ref{ssect:freefermion}). This means that we are interested in the following two-point
function 
\begin{equation}
\langle T^{\alpha}(0,0)\bar{T}^{\alpha}(x,0)\rangle_{\rho_w^{\otimes\alpha}}\ .
\end{equation}
From the BFT perspective, this is obtained by focusing on the
\emph{purely spatial direction}, namely, we consider an ``horizontal
path'' by setting $\gamma=\pi/2$ in (\ref{eq:largedev}) (and we take $h(\theta) =h_p$). Each two-point
function of $U(1)$ twist fields in (\ref{eq:U1_twist}) reads
\begin{equation}\label{eq:tautauspace}
\langle\tau_{p}^{\alpha}(0,0)\bar{\tau}_{p}^{\alpha}(x,0)\rangle_{\rho_w}\asymp\exp\left\{ x\:F_p(-\ii)\right\} ,\quad F_p(-\ii)=\int\frac{d\theta}{2\pi}\log\left(\frac{1+e^{\ii h_{p}-w(\theta)}}{1+e^{-w(\theta)}}\right)\ .
\end{equation}
Then we consider the product in Eq. (\ref{eq:U1_twist}), which turns into a sum in the exponent, i.e.,
\begin{equation} \label{F_alpha}
\langle T^{\alpha}(0,0)\bar{T}^{\alpha}(x,0)\rangle_{\rho_w^{\otimes \alpha}}\asymp\exp\left\{x\:F^{\alpha}(-\ii)\right\},
\quad 
%
{\rm with } \quad 
%
    F^{\alpha}(-\ii) = 
        \sum_{q=-\alpha/2+1}^{\alpha/2} F_{2q-1} (-\ii)\ .
\end{equation}
We may further evaluate those sums, by considering separately the part which depends and the part which does not depend on $p$ (equivalently $q,q'$, Eq.~(\ref{eq:decomposition})).
The latter is trivial and simply gives a contribution to
$F^{\alpha}(-\ii)$ which is $-\int d\theta/(2\pi)$ of
\begin{equation}
2\sum_{q=1}^{\alpha/2} 
\log\left(1+e^{-w(\theta)}\right)=\alpha\log\left(1+e^{-w(\theta)}\right)\ .
\end{equation}
For the remaining part, let us start by focusing
on half of the sum, the terms from $q=1$ to $\alpha/2$, in (\ref{F_alpha}). By
defining $z=\frac{2\pi \ii}{\alpha}$, $s=w+\frac{\pi \ii}{\alpha}$,
we get 
\begin{eqnarray} \label{step1}
\sum_{q=1}^{\alpha/2}\log(1+e^{zq-s}) & = & \sum_{r=1}^{\infty}\frac{(-1)^{r+1}}{r}e^{-r(s-z)}\left(\frac{1-e^{rz\alpha/2}}{1-e^{rz}}\right)\label{eq:zs}
\end{eqnarray}
where we used the Taylor expansion $\log(1+x)=\sum_{r=1}^{\infty}(-1)^{r+1}x^{r}/r$ (which converges for $w>0$),
and we performed the sum over $q$. Next, we want to perform the sum
in $r$ in the r.h.s.~of \eqref{step1}. To do that, we substitute the values of $z$ and $w$ first:
\begin{equation}
\sum_{r=1}^{\infty}\frac{(-1)^{r+1}}{r}e^{-r(w-\frac{\pi \ii}{\alpha})}\left(\frac{1-e^{r\pi \ii}}{1-e^{r\frac{2\pi 
\ii}{\alpha}}}\right)\label{eq:sumr}
\end{equation}
where now we should consider separately
three cases:
\begin{enumerate}
\item $r=\alpha m$ for integer $m$: in this case $r$ is even (as $\alpha$ is even), and we have
\begin{align}
\sum_{m=1}^{\infty}\frac{(-1)^{\alpha m+1}}{\alpha m}e^{-\alpha mw+m\pi i}\left(\frac{r\pi i}{2\pi ir/\alpha}\right) & =\sum_{m=1}^{\infty}\frac{(-1)^{m+1}}{2 m}e^{-\alpha mw}\\
 & =\frac{1}{2}\log\left(1+e^{-\alpha w}\right)\ .
\end{align}
\item $r$ even but $r\neq\alpha m$ for any integer $m$: in this case each term of the sum
(\ref{eq:sumr}) is zero due to the vanishing of the numerator, i.e.,
$(1-e^{r\pi \ii})=0$.
\item $r$ odd: this gives
\begin{equation}
\sum_{r\:\text{odd}}\frac{2}{r}e^{-rw}\left(\frac{e^{r\frac{\pi \ii}\alpha}}{1-e^{r\frac{2\pi \ii}{\alpha}}}\right)=\sum_{r\:\text{odd}}\frac{\ii}{r}\frac{e^{-rw}}{\sin\frac{\pi r}{\alpha}}\ .\label{eq:sumr-1}
\end{equation}
\end{enumerate}
The sum of the terms for $q=-\alpha/2+1$ to $0$ in (\ref{F_alpha}) give exactly the complex conjugate of this result. Thus we get
\begin{equation}
    \sum_{q=-\alpha/2+1}^{\alpha/2}\log(1+e^{zq-s})
    = \log(1+e^{-\alpha w})\ .
\end{equation}
Putting everything together, $F^{\alpha}(-\ii)$ in \eqref{F_alpha} can be written as
\begin{equation}
F^{\alpha}(-i)=\int\frac{d\theta}{2\pi}\left[\log\left(1+e^{-\alpha w(\theta)}\right)-\alpha\log\left(1+e^{-w(\theta)}\right)\right]\ .
\end{equation}
Finally, it is a matter of simple algebra to show that, in terms of
the occupation function $n(\theta)$ \eqref{eq:occupation}, we get
\begin{equation}
F^{\alpha}(-i)=\int\frac{d\theta}{2\pi}H^{\alpha}(\theta)
\end{equation}
where we defined
\begin{eqnarray}
H^{\alpha}(\theta) & = & \frac{1}{1-\alpha}\log\left[n(\theta)^{\alpha}+(1-n(\theta))^{\alpha}\right]\ .\label{eq:Halpha}
\end{eqnarray}
The $\alpha-$R\'enyi entropy is finally given by 
\begin{equation}
S_{\alpha}(x)=\frac{1}{1-\alpha}\log\langle T^{\alpha}(x,0)\bar{T}^{\alpha}(0,0)\rangle_{\rho_w^{\otimes \alpha}}\sim x\:\int\frac{d\theta}{2\pi}H^{\alpha}(\theta)\ ,\label{eq:Renyi_purespace}
\end{equation}
which coincides with the results obtained in \cite{alba2017quench,mestyan2018renyi} (there in the more general context of interacting integrable models).

\subsection{Long-range correlations due to correlated particle pairs in homogeneous global quenches} \label{ssect:longrange}

We now review the main concepts underlying quantum quenches, restricting to ``integrable" pair-producing initial states, and we explain how long-range correlations develop after such quenches.

A quantum quench is an initial value problem for the many-body system where the initial state is the ground state of a different Hamiltonian than that used for the time evolution. Typically, one imagines a sudden change of parameter, for instance of the mass parameter. In integrable models, certain quenches are known to be of ``integrable" type \cite{piroli2017integrable,delfino2014quantum,delfino2017theory}. In these cases, the initial state can be written explicitly in terms of the scattering states (or Bethe ansatz states) of the post-quench, evolution Hamiltonian, as a so-called ``squeezed state":
\begin{equation} \label{pair_state}
\ket{\Psi} = \frac1{\mathcal N}\exp\left(\frac12 \int d\theta\,\mathcal{K}_{\theta,-\theta}\psi^\dag_\theta \psi^\dag_{-\theta}\right)\ket{0}
\end{equation}
for some ($\theta$-dependent) factor $ \mathcal{K}_{\theta, -\theta},$ with $\mathcal N_\theta$ denoting a normalization constant, and $\left|0\right\rangle $ being the ground state of the post-quench Hamiltonian. The squeezed state is generically a finite-density state, where the energy (of the post-quench Hamiltonian) is extensive with the system size. See App.~\ref{ssect:quench} for a discussion of such integrable initial states in free fermion models. We will use later the fact there is a Bogolioubov transformation of the fermionic mode operators (a transformation between the post-quench and pre-quench fermions),
\begin{equation}\label{eq:bogoabstract}
    \psi(x,t) \leftrightarrow \tilde\psi(x,t)\qquad\mbox{(Bogolioubov)}
\end{equation}
such that the squeezed state satisfies (is defined by)
\begin{equation}\label{eq:initzero}
    \tilde\psi(x,t)|\Psi\rangle = 0.
\end{equation}

After a long time in a quench problem, the state locally approaches a GGE. In integrable quenches, there is a well-known relation between the squeezed-state representation of the initial state, and the long-time GGE (see e.g.~\cite{Essler_2016}). The statement of convergence to a GGE pertains only to local operators, or operators supported on finite intervals (that do not grow with time):
\begin{equation}\label{eq:thermalisation}
    \langle \Psi|a(x,t)|\Psi\rangle \to \langle a(x)\rangle_{\rho_w},\quad  t\to\infty\ .
\end{equation}
The limit in \eqref{eq:thermalisation} is expected to be valid everywhere in space.
The relation between initial state and long-time GGE in free fermions can be worked out explicitly (see \eqref{eq:Kwapp})
\begin{equation}\label{eq:quenchgge}
    e^{-w(\theta)} = |\mathcal K_{\theta,-\theta}|^2\ .
\end{equation}
Namely, we see that the map from squeezed states to GGEs is in fact one-to-one.

Because of this one-to-one correspondence, it is clear that it is sufficient to know the long-time GGE in order to know the full behaviour of correlation functions in spacetime, as the GGE fixes the initial state uniquely (naturally under the condition that it be a pair-producing squeezed state). However, this relation can be relatively complex. Indeed, the statement of generalised thermalisation -- that a GGE is reached -- is true, generically, only on finite regions of space (see App.~\ref{app:approachGGE}). As is typically the case out of equilibrium, on large regions, say regions that grow linearly with the time after the quench, the state might not correctly be described by a GGE; and this may even be true for all times! Instead, the state may admit long-range spatial correlations, for instance correlations that have a large weight on distances that grow linearly with the time. These are not present in GGEs; recall that, as discussed above, GGEs typically have correlations that decay quickly enough in space (see App.~\ref{app:global}). Thus, even at long times, there may remain effects of the initial state that are not described by a GGE.

In the case of a squeezed state, such long-range correlations indeed exist, as we show in App.~\ref{app:approachGGE} (and also in App.~\ref{ssect:decay} for ``single-mode densities and currents", introduced below in Sec.~\ref{ssect:singlemode}). Their interpretation is that they are due to production of correlated pairs of opposite-momentum particles by the quench protocol. These particle pairs carry correlations to large distances as they separate. We evaluate explicitly these long-range correlations for conserved densities and currents in App.~\ref{app:approachGGE} (and App.~\ref{ssect:decay}). For instance, we find, for the charge density $q(x,t) = \psi^\dag(x,t)\psi(x,t)$, that $\langle \Psi|q(x,t)q(x',t)|\Psi\rangle - \langle \Psi|q(x,t)|\Psi\rangle \langle\Psi| q(x',t)|\Psi\rangle$ exhibits strong, ballistic-scale correlations, in accordance with the picture according to which particle pairs are emitted at all velocities admitted by the dispersion relation.

In order to evaluate the R\'enyi entanglement entropy, as is clear from the calculation for GGEs in Sec.~\ref{subsec:spacedirection}, we must evaluate the large-deviation theory for fluctuations of charges on large regions of space, and / or, as we will see below, fluctuations of current on large intervals of time. The BFT allows us to do that. However, as mentioned, the BFT requires no long-range correlations along the path $\underline\ell$ in \eqref{eq:Delta_j}, as the flow equation (cf. \eqref{eq:flow}) assumes that the state along the path is a GGE. Long-range correlations may break the assumption that scaled cumulants are evaluated within a GGE -- the effects of long-range correlations on the 2nd cumulants is illustrated in App.~\ref{app:approachGGE}. Below, we take them into account by {\em choosing appropriately the path $\underline\ell$ in order to avoid such correlations}! Thus, the knowledge of {\em where} such correlations exist, and the knowledge of the long-time GGE, is sufficient.

The fact that there exists a path $\underline\ell$ that avoids long-range correlations explains why, in pair-production states, the full behaviour of the R\'enyi entropies can be written in a simple and universal way in terms of the long-time GGE. 
The same is not true when considering non-integrable quenches, namely quenches from more complicated states where groups of more than two correlated particles are emitted. Without the constraint of pair-production, the one-to-one correspondence between the state and the GGE is lost, and long-range correlations carry additional information not present in the GGE. Then, from our approach, we see that a universal description in terms of the long-time GGE is lost because such an ``avoiding path'' does not exist in general anymore.

\subsection{R\'enyi entropies of half system after a quench (current fluctuations in time)} \label{ssect:renyihalf}

We now turn to the calculation of the $\alpha-$R\'enyi entropy of
a semi-infinite interval $A=[0,\infty)$ after a global homogeneous quantum quench, at long times $t\to\infty$.  This is obtained
from the branch-point twist field one-point function
\begin{equation}
\langle \Psi^\alpha|T^{\alpha}(0,t)| \Psi^\alpha\rangle
\end{equation}
in the state $| \Psi^\alpha\rangle = |\Psi\rangle^{\otimes \alpha} = \prod_{i=1}^\alpha |\Psi_i\rangle$, the $\alpha$-copy replica of \eqref{pair_state},
\begin{equation}\label{eq:psi_theta_i} 
\left|\Psi^{\alpha}\right\rangle =\prod_{i=1}^{\alpha}
\frac1{\mathcal N}\exp\left(\frac12 \int d\theta\,\mathcal{K}_{\theta,-\theta}\psi^\dag_{\theta,i} \psi^\dag_{-\theta,i}\right)\ket{0,i}\ .
\end{equation}

\begin{figure}
    \centering
	\includegraphics[width=\textwidth]{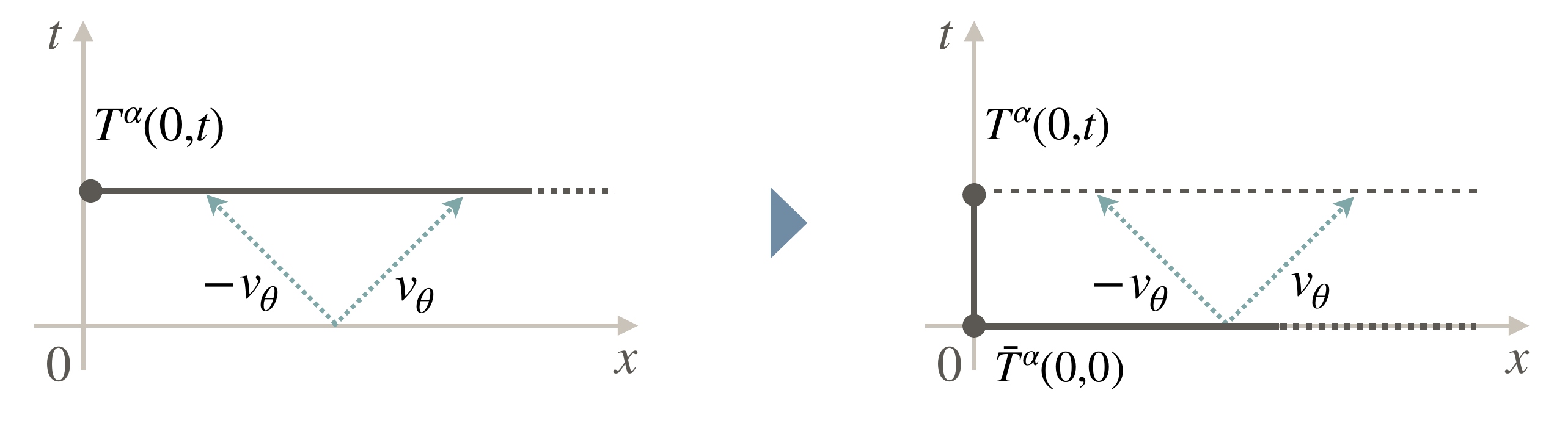}
	\caption{\label{quench_halfsystem}Evolution of R\'enyi entropies of half system $A=[0,\infty]$ within BFT. Left: Initial integration path. Because of initially entangled pairs, points along this path at time $t$ will be correlated, which prevents us from applying BFT directly. Right: Deformed integration path. Along this new path one can show that point are not correlated anymore. Moreover the only term contributing to the growth in time of entanglement is the vertical path from $(0,t)$ to $(0,0)$.}  
\end{figure}

As expressed in \eqref{eq:thermalisation}, one-point functions of local observables converge to averages within GGEs. However, as we discussed, twist-fields are ``semi-local" observables; from the point $(0,t)$ emanates a branch cut, which is sensitive to the state where it passes. The branch cut can be taken on the horizontal half-line $\{(x,t):x\in[0,\infty)\}$ going from $(0,t)$ to $(\infty,t)$, as done in the explicit construction of the field in Sec.~\ref{sec:entanglement_twist}. As explained in Sec.~\ref{ssect:longrange}, and analysed in App.~\ref{app:approachGGE}, along this half-line, there exist long-range correlations due to coherent particle pairs emitted by the initial state. This prevents us from applying the BFT along this path (see Fig.\ref{quench_halfsystem} (left)).

Using path independence of twist fields correlation functions, we can deform the path, between its initial and final points, in a way to avoid such correlations. Specifically, we choose the piece-wise linear
path joining the points $(0,t)\to(0,0)\to(\infty,0)$.
This is shown in Fig.~\ref{quench_halfsystem} (right). We note that as the final point is at spatial infinity, it can be displaced to time 0 -- this in fact implements the correct physics of the entanglement entropy due to the single boundary at $x=0$. Then, we may represent the one-point function as
\begin{equation}\label{eq:onethree}
\langle \Psi^\alpha|T^{\alpha}(0,t)| \Psi^\alpha\rangle \asymp
\langle \Psi^\alpha|T^{\alpha}(0,t)\bar T^\alpha(0,0) T^\alpha(0^+,0)| \Psi^\alpha\rangle
\end{equation}
where the factors $T^{\alpha}(0,t)\bar T^\alpha(0,0)$ represent the segment of path $(0,t)\to(0,0)$, and the factor $T^\alpha(0^+,0)$, the segment $(0,0)\to(\infty,0)$. This is valid as an asymptotic relation for large $t$, where the UV singularity due to the proximity of the fields $\bar T^\alpha(0,0)$ and $T^\alpha(0^+,0)$ (which occurs because of renormalisation effects) is neglected.

We simplify the expression \eqref{eq:onethree} in two steps.

First, we note that the segment of path $(0,0)\to(\infty,0)$ does not provide any contribution to the result. This is because we may re-write the branch-point twist field $T^\alpha(0^+,0)$ as is done in Sec.~\ref{sec:entanglement_twist}, but {\em in the basis of the before-quench canonical free fermions of the replica theory}, $\tilde\psi_i(x,0)$, Eq.~\eqref{eq:bogoabstract}. The exchange relations \eqref{eq:permexch1} (here at $t=0$) hold for any field $a_i(x,0)$, and in particular hold for $a_i(x,0) = \tilde\psi_i(x,0)$. Therefore, by the same arguments, we obtain a decomposition as in \eqref{eq:decomposition},
\begin{equation}
    T^{\alpha}=\prod_{p\in I_{\alpha}}\tilde\tau_{p}^{\alpha}
\end{equation}
but for {\em different} $U(1)$ twist fields
\begin{equation}
    \tilde\tau_p^\alpha(x,0)
    = \exp\Big[
    \ii \int_{x}^{\infty} dx'\,
    \tilde q_p(x',0)
    \Big],\qquad \tilde q_p(x,0) = \frac{\pi p}\alpha \tilde \psi_p^\dag(x,0)\tilde\psi_p(x,0)
\end{equation}
instead of \eqref{eq:tauexp}. By \eqref{eq:initzero}, we have $\tilde\psi_i(x,0)|\Psi_j\rangle = 0$ for all $i,j$, so it is clear that
\begin{equation}
    \tilde q_p(x,0)|\Psi^\alpha\rangle = 0\ ,
\end{equation}
therefore
\begin{equation}
        \tilde\tau_p^\alpha(x,0)|\Psi^\alpha\rangle = |\Psi^\alpha\rangle\ .
\end{equation}
Hence
\begin{equation}
\langle \Psi^\alpha|T^{\alpha}(0,t)| \Psi^\alpha\rangle \asymp
\langle \Psi^\alpha|T^{\alpha}(0,t)\bar T^\alpha(0,0)| \Psi^\alpha\rangle\ .
\end{equation}

Second, we note that along the path $(0,t)\to(0,0)$, generic observables do not have long-range correlations coming from pair productions: correlations of generic observables approach those within the final GGE fast enough, in such a way that corrections due to the quench give only {\em sublinear corrections to cumulants of time-integrated fermion bilinears} (such as conserved densities and currents). This is because particle pairs always create correlations between points at separate spatial coordinates: it is not possible to create two co-propagating fermions, with the same momentum (here they would be with vanishing momentum, as the total momentum has to be zero). This fact is discussed in App.~\ref{app:approachGGE}; the discussion there is for a single copy, but it extends immediately to the $\alpha$-copy state $\ket{\Psi^\alpha}$. Therefore, rewriting the branch-point twist fields in terms of $U(1)$ $\tau_p^\alpha$ with branches in the time direction, using \eqref{eq:decomposition}, and expanding in cumulants of $U(1)$ currents, we see that  on the segment $(0,t)\to(0,0)$, for the purpose of the BFT, the state is correctly described by a GGE\footnote{We remark that for bosonic system, this argument would break, as pairs of particles with equal, zero momenta are emitted with a finite density. However, it turns out that this correction due to the quench would not affect the cumulants of time-integrated currents, as such pairs, being of zero momentum, do not carry any current.}.

An important consequence of these arguments is that {\em we may evaluate the twist field one-point function after a quench, as an equal-space, different-time two-point function within the GGE representing the final state, Eq.~\eqref{eq:quenchgge}}: \begin{equation}\label{eq:quenchtimegge}
\langle \Psi^\alpha|T^{\alpha}(0,t)| \Psi^\alpha\rangle \asymp \langle T^{\alpha}(0,t)\bar{T}^{\alpha}(0,0)\rangle_{\rho_w^{\otimes \alpha}}\ .
\end{equation}
This is valid at long times, and omits small-time effects that occur before generalised thermalisation (which do not affect the asymptotic regime we look at).

In order to apply BFT to the r.h.s. of \eqref{eq:quenchtimegge} a last observation is needed. As the GGE is a Wick-theorem state, we can use \eqref{eq:U1_twist}, thus we are interested in the separate two-point functions of $U(1)$ twist fields $\tau_p^\alpha$ with branches in the time direction. It turns out that, as emphasised in Sec.~\ref{subsec:Flambda}, the currents $j_p(0,t'),\;t'\in[0,t]$ in GGEs have time-correlations that decay fast enough so as to give only linearly growing cumulants: this is what allows the application of the BFT (see App.~\ref{app:global} for a full discussion). We note that this is not true in general of other observables: in GGEs, generic fermion bilinears have cumulants that grow faster than linearly with time. But we are intersted in the currents only.

We are now in position to apply standard BFT. This amounts to repeating the same calculation as above in Sec.~\ref{subsec:spacedirection},
but now in the \emph{purely temporal direction}. We use the general formula
(\ref{eq:largedev}) for the ``vertical'' path connecting
initial and final point by choosing $\gamma=0$, and similarly get (note that the path is in opposite direction as that of formula
(\ref{eq:largedev}), and thus we must take $h(\theta) = -h_p$)
\begin{align}
\langle\tau_{p}^{\alpha}(0,t)\bar{\tau}_{p}^{\alpha}(0,0)\rangle_{\rho_w} & \asymp\exp\left\{ t\:F_p(-\ii)\right\} ,\quad F_p(-\ii)=\int\frac{d\theta}{2\pi}|v(\theta)|\log\left(\frac{1+e^{\ii h_p \sgn(\theta)-w(\theta)}}{1+e^{-w(\theta)}}\right)
\end{align}
where we used $\sgn(v(\theta)) = \sgn(\theta)$.
Again, after performing the product all two-points functions of $U(1)$ twist
fields, we get 
\begin{align}
\langle T^{\alpha}(0,t)\bar{T}^{\alpha}(0,0)\rangle_{\rho_w^{\otimes\alpha}}& \asymp\exp\left\{ t\:F^{\alpha}(-\ii)\right\} ,\quad F^{\alpha}(-\ii) = \int\frac{d\theta}{2\pi}|v(\theta)|\log\left[\frac{1+e^{-\alpha w(\theta)}}{\left(1+e^{-w(\theta)}\right)^\alpha}\right]\ .
\end{align}
Using $H_{\alpha}(\theta)$ as defined in (\ref{eq:Halpha}), the
$\alpha-$R\'enyi entropy reads
\begin{equation}
S_{\alpha}(t)=\frac{1}{1-\alpha}\log\langle T^{\alpha}(0,t)\bar{T}^{\alpha}(0,0)\rangle_{\rho_w^{\otimes\alpha}} \sim t\int\frac{d\theta}{2\pi}|v(\theta)|H_{\alpha}(\theta)\ .\label{eq:Renyi_puretime}
\end{equation}
This is the result obtained both from exact calculation in \cite{fagotti2008evolution}
and within the quasi-particle picture in \cite{alba2017entanglement,alba2018entanglementlong}.


\subsection{Single-mode and pair-mode twist fields}
\label{ssect:singlemode}

We have discussed in Sec.~\ref{ssect:freefermion} the conserved quantities $Q_\theta = \psi^\dagger_\theta\psi_\theta$, forming a ``scattering" or continuous basis for the extensive conserved quantities of the free fermion model. In Sec.~\ref{ssect:enhanced}, we discussed the replica model with $\alpha$ copies, and the $U(1)$ charges $Q_p$, which are just the integration $Q_p = h_p \int d\theta\,Q_{\theta,p}$ (with $h_p=\frac{\pi p}\alpha$) over all momenta $\theta$ of the continuous basis $Q_{\theta,p} = \psi^\dagger_{\theta,p} \psi_{\theta,p}$ in the Fourier-copy $p$. There, we have also discussed the twist fields $\tau_p^\alpha$ associated to these charges, which turned out to be useful in the computation of the R\'enyi entanglement entropies in Subsections \ref{subsec:spacedirection} and \ref{ssect:renyihalf}. A natural extension of these constructions is to the twist fields associated to each conserved quantity $Q_{\theta,p}$. As we will see, these are indeed useful in evaluating the behaviour of R\'enyi entanglement entropies for intervals that grow linearly with time.

In order to simplify the notation, we consider a single copy of the fermion, and the scattering basis $Q_\theta$; the discussion immediately adapts to the Fourier-copy $p$.

In the study of the ballistic behaviours of many-body systems, and in particular in the BFT, it is essential that the conserved charge $Q$ considered be extensive -- scale linearly with the volume (typically one requires $\braket{Q^2}^c \propto L$ \cite{ilievski2016quasilocal,doyonpseudo}). The charges $Q_\theta$ are not extensive. However, as they form a continuous basis, integrals on small $\theta$-intervals are extensive; thus it is better to define, for $\epsilon>0$ as small as desired,
\begin{equation}
    Q_\theta = \int_{\theta-\epsilon/2}^{\theta+\epsilon/2}
    d\theta'\,\psi_{\theta'}^\dagger \psi_{\theta'}\ .
\end{equation}
These act as
\begin{equation}
    Q_{\theta} \ket{\theta'}
    = \Theta(\epsilon/2 - |\theta'-\theta|)\ket{\theta'}
\end{equation}
hence have one-particle eigenvalues
\begin{equation}\label{eq:hsingle}
    h_{\theta} (\theta') = \Theta(\epsilon/2 - |\theta'-\theta|).
\end{equation}
We show in App.~\ref{app:singlemode} that such $Q_\theta$ are indeed extensive in GGEs, and we evaluate explicitly their associated densities and currents $q_\theta(x,t)$ and $j_\theta(x,t)$,
\begin{equation}
    Q_\theta = \int dx\,q_\theta(x,t),\quad
    \partial_t q_\theta(x,t) + \partial_x j_\theta(x,t) = 0\ .
\end{equation}
From this, one can immediately construct the associated twist field
\begin{equation}
    \boldsymbol \tau_\theta(x,t)
    = \exp \Big[\ii \int_x^\infty dx'\,
    q_\theta(x',t)
    \Big]
\end{equation}
and, for its correlation functions, apply the corresponding BFT based on the one-particle eigenvalue \eqref{eq:hsingle}.

In fact, we are interested in studying the squeezed state \eqref{pair_state}. It is clear that this state factorises into momentum intervals as follows:
\begin{equation}
    \ket\Psi = \prod_{\theta \in (\mathbb N+ \frac12)\epsilon} \ket{\Psi_{|\theta|}}
\end{equation}
where
\begin{equation}
\ket{\Psi_{|\theta|}} = \frac1{\mathcal N_{|\theta|}}\exp\left(\int_{\theta-\epsilon/2}^{\theta+\epsilon/2} d\theta'\,\mathcal{K}_{\theta,-\theta}\psi^\dag_\theta \psi^\dag_{-\theta}\right)\ket{0_{|\theta|}}
\end{equation}
and we write the ground state in a naturally factorised way as $\ket0 = \prod_{\theta\in(\mathbb N+\frac12)\epsilon} \ket{0_{|\theta|}}$. Likewise, we will consider the pair-mode charges $Q_{|\theta|} = Q_\theta + Q_{-\theta}$ and the associated densities
\begin{equation}
    q_{|\theta|}(x,t) = q_\theta(x,t) + q_{-\theta}(x,t)\ .
\end{equation}
Both act trivially (as zero) on $|\Psi_{|\theta'|}\rangle$ if $\theta'\neq \theta$ ($\theta,\theta' \in (\mathbb N+\frac12)\epsilon$). From these, we get the pair-mode twist fields
\begin{equation}
    \boldsymbol \tau_{|\theta|}(x,t)
    = \exp \Big[\ii \int_x^\infty dx'\,
    q_{|\theta|}(x',t)
    \Big],
\end{equation}
which acts trivially (as the identity) on $|\Psi_{|\theta'|}\rangle$ if $\theta'\neq \theta$.

These are still $U(1)$ twist fields, for the sub-$U(1)$ symmetry acting on the tensor factor of modes within $[\theta-\epsilon/2,\theta+\epsilon/2]$. Note in particular that the global $U(1)$ twist field $\tau(x,t)$ associated to the total charge $Q = \int d\theta\,\psi^\dagger_\theta \psi_\theta = \int dx\,\psi^\dagger(x)\psi(x)$ can be factorised as
\begin{equation}\label{eq:taufactsingle}
    \tau(x,t) = \prod_{\theta \in (\mathbb N+ \frac12)\epsilon}
    \boldsymbol \tau_{|\theta|}(x,t)
\end{equation}
and that, by factorisation of its action on the state, we have
\begin{equation}
    \bra\Psi \tau(x,t)\tau(x',t')\ket\Psi
    = \prod_{\theta \in (\mathbb N+ \frac12)\epsilon}
    \bra{\Psi_{|\theta|}}
    \boldsymbol \tau_{|\theta|}(x,t)
    \boldsymbol \tau_{|\theta|}(x',t')
    \ket{\Psi_{|\theta|}}.
\end{equation}
Clearly, as the pair-mode twist fields act trivially on other tensor factors in the state, we may also write, more simply,
\begin{equation}
\label{eq:taufactsingletwopoint}
    \bra\Psi \tau(x,t)\tau(x',t')\ket\Psi
    = \prod_{\theta \in (\mathbb N+ \frac12)\epsilon}
    \bra{\Psi}
    \boldsymbol \tau_{|\theta|}(x,t)
    \boldsymbol \tau_{|\theta|}(x',t')
    \ket{\Psi}.
\end{equation}

\subsection{R\'enyi entropies of an interval after a quench (fluctuations of \emph{single-mode} densities and currents)} \label{ssect:renyisingle}

We finally extend the result for the entanglement growth after a quench to a finite, but ballistically growing interval
$A=[0,x]$, with $x=\xi t$. To this aim, we should consider the following two-point correlation function in a squeezed state Eq.~\eqref{pair_state} (or Eq.~\eqref{eq:psi_theta_i} in the replicated theory):
\begin{equation}
\langle\Psi^\alpha| T^{\alpha}(0,t)\bar{T}^{\alpha}(x,t)|\Psi^\alpha\rangle,\quad
x = \xi t,\quad t\to\infty.
\end{equation}
The idea is the same as that used in Sec.~\ref{ssect:renyihalf}, that we need to deform the integration path in such a way that, everywhere along the path, all points remain uncorrelated (on large scales), thus enabling us to apply BFT. The choice of the path will now depend on the values of $\xi$, and in fact, we will need re-write the two-point function as a product of two-point functions of pair-mode twists fields, and to choose different paths for each such two-point function.

It will simplify the discussion to already re-write the two-point function in terms of $U(1)$ twist field. As the squeezed state is a Wick-theorem state, we can directly use \eqref{eq:U1_twist}:
\begin{equation}
\langle\Psi^\alpha| T^{\alpha}(0,t)\bar{T}^{\alpha}(x,t)|\Psi^\alpha\rangle=\prod_{q=-\alpha/2+1}^{\alpha/2}\langle\Psi_{2q-1}|\tau_{2q-1}^{\alpha}(0,t)\bar{\tau}_{2q-1}^{\alpha}(x,t)|\Psi_{2q-1}\rangle\label{eq:U1_twist_psi}
\end{equation}
where $|\Psi_p\rangle$ is the squeezed state $|\Psi\rangle$ for the fermions $\psi_p(x),\psi^\dagger_p(x)$ on Fourier-copy space $p$. 

\begin{figure}
    \centering
	\includegraphics[width=0.85\textwidth]{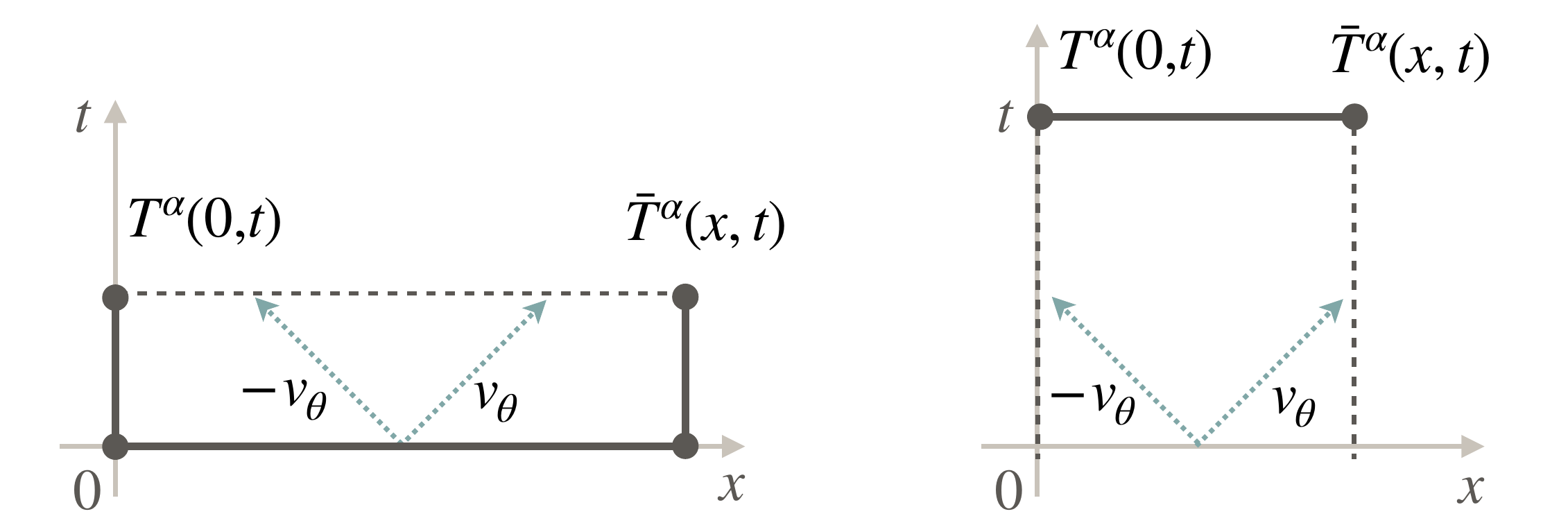}
	\caption{\label{quench_finitesystem}Evolution of R\'enyi entropies of finite subsystem $A=[0,x]$ within BFT. The integration path that we need to chose (continuous dark-gray line) in order for BFT to apply is different at short (left) and long (right) times. 
    The choice depends on which points in spacetime get correlated because of initially entangled pairs produced by the initial state.}  
\end{figure}

We start by considering the two asymptotic regimes:
\begin{itemize}
\item At short times (more precisely in the limit $\xi \to \infty$ of the scaled, long-time asymptotic behaviour), entangled particle pairs coming out from
the initial state will correlate points within the original integration path. To apply BFT, then, we deform the straight path $(0,t)\to(x,t)$ to the piece-wise straight path  $(0,t)\to(0,0)\to(x,0)\to(x,t)$, made of three segments
(see Fig.~\ref{quench_finitesystem} (left)). By the same arguments as in Sec.~\ref{ssect:renyihalf}, the space-like segment will
not contribute to the entanglement growth, and the time-like segments will give separated, identical contributions given by the long-time GGE. We are thus left with the contribution of the two, separate time-like segments. The fact that the segments do not correlate with each other is thanks to the assumption that the GGE satisfies $n(\theta)\to0$ as $|\theta|\to\infty$ (that is, the density of pairs produced at large momenta tends to zero), as is discussed in App.~\ref{app:approachGGE}.
\item At long enough times (either the limit $\xi \to 0$ of the scaled, long-time asymptotic behaviour, or the long-time limit followed by the long-distance scaling), the particles generated from the initial state do not correlate points within the path $(0,t)\to (x,t)$: cumulants scaled by the distance $x$ do not receive contributions from such particle pairs. Corrections terms to the GGE values of  cumulants can only come from pairs of particles at infinitesimally small momenta, and, it turns out, such corrections become zero when the total number of correlated pairs on the interval $[0,x]$ tend to zero. As there is at most a finite density of pairs produced per unit momenta, there remain no pairs on infinitesimally small momentum intervals\footnote{In fact, as we are looking at fermionic models, the density tends to zero at zero momenta, but this is not required in this argument.}. Thus the asymptotic behaviour is that obtained from the long-time GGE. This is discussed in App.~\ref{app:approachGGE}. Particle pairs of finite momenta would, of course, correlate points between the path segments $(0,t)\to(0,0)$ and $(x,0)\to(x,t)$ (also discussed in App.~\ref{app:approachGGE}), thus we must avoid the piece-wise straight path. Therefore the correct way to use BFT is by using the original path
(see Fig.~\ref{quench_finitesystem} (right).
\end{itemize}

So we arrive to the following asymptotic results for $x,t\to\infty$:
\begin{eqnarray}
\langle\Psi^\alpha| T^{\alpha}(0,t)\bar{T}^{\alpha}(x,t|\Psi^\alpha\rangle & \asymp & \begin{cases}
\langle T^{\alpha}(0,t)\bar{T}^{\alpha}(0,0)\rangle_{\rho_w}\langle  T^{\alpha}(x,0)\bar{T}^{\alpha}(x,t)\rangle_{\rho_w} & t\ll x\\
\langle T^{\alpha}(0,t)\bar{T}^{\alpha}(x,t)\rangle_{\rho_w} & t\gg x
\end{cases}\nonumber \\
 & = & \begin{cases}
\left| \langle T^{\alpha}(0,t)\bar{T}^{\alpha}(0,0)\rangle_{\rho_w} \right|^2 & t\ll x\\
\langle T^{\alpha}(0,t)\bar{T}^{\alpha}(x,t)\rangle_{\rho_w} & t\gg x
\end{cases}
\label{eq:factorisation_regimes}
\end{eqnarray}
where we used $\langle T^{\alpha}(0,t)\bar{T}^{\alpha}(0,0)\rangle_{\rho_w}^* = \langle T^{\alpha}(0,0)\bar{T}^{\alpha}(0,t)\rangle_{\rho_w} = \langle T^{\alpha}(x,0)\bar{T}^{\alpha}(x,t)\rangle_{\rho_w}$ (by the fact that $T(x,t)^\dagger = \bar T(x,t)$). This
leads to
\begin{eqnarray}
S_{\alpha}(x,t)=\frac{1}{1-\alpha}\log\langle T^{\alpha}(0,t)\bar{T}^{\alpha}(x,t)\rangle & \sim & \begin{cases}
\displaystyle 2t\int\frac{d\theta}{2\pi}|v(\theta)|H_{\alpha}(\theta) & t\ll x\\[0.5cm]
\displaystyle 
x\int\frac{d\theta}{2\pi}H_{\alpha}(\theta) & t\gg x.
\end{cases}\label{eq:asymptotics}
\end{eqnarray}
Namely, at short times (but much larger than microscopic times), the growth is described by the path in the purely temporal direction (\ref{eq:Renyi_puretime}), and at long times, the system goes, uniformly as a function of the velocity $x/t\to0$, to the equilibrium GGE and
there the result is the one of the purely spatial path (\ref{eq:Renyi_purespace}). 

These are, however, only asymptotic results in $\xi$, within the scaled regime $x,t\propto \ell$. It turns out that we can access all values of $\xi=x/t$ within this regime, by using similar arguments, but now for the {\em single-mode twist fields} introduced in Sec.~\ref{ssect:singlemode} (in fact, we need the pair-mode twist fields). Effectively, using these, we will be able to take into account that the meaning of ``short'' and ``long'' time depend directly on the speed of the travelling particles
$v(\theta) = E'(\theta)$.

We start with the decomposition of global $U(1)$ twist fields into pair-mode twist fields \eqref{eq:taufactsingle}, which we write in the replica model for each Fourier-copy $p$, and with single-particle charge eigenvalue $h_p = \frac{\pi p}\alpha$ instead of $1$ in \eqref{eq:taufactsingle} (as done in \eqref{eq:tauexp}) 
\begin{equation}
    \tau^\alpha_p(x,t) = \prod_{\theta \in (\mathbb N+ \frac12)\epsilon}
    \boldsymbol \tau_{|\theta|,p}^\alpha(x,t)
\end{equation}
where
\begin{equation}
    \boldsymbol \tau_{|\theta|,p}^\alpha(x,t)
    = \exp \Big[\ii \int_x^\infty dx'\,
    q_{|\theta|,p}(x',t)
    \Big]
\end{equation}
and $q_{|\theta|,p}(x,t)$ has the form \eqref{eq:appendixA:single_mode_density} times $h_p$. Thus, by factorisation of two-point functions \eqref{eq:taufactsingletwopoint}, we re-write \eqref{eq:U1_twist_psi} as
\begin{equation}
\langle\Psi^\alpha| T^{\alpha}(0,t)\bar{T}^{\alpha}(x,t)|\Psi^\alpha\rangle=\prod_{\theta \in (\mathbb N+ \frac12)\epsilon}\prod_{q=-\alpha/2+1}^{\alpha/2}\langle\Psi_{2q-1}|\boldsymbol\tau_{|\theta|,2q-1}^{\alpha}(0,t)\bar{\boldsymbol\tau}_{|\theta|,2q-1}^{\alpha}(x,t)|\Psi_{2q-1}\rangle\ .\label{eq:U1_twist_psi_single}
\end{equation}

Having made this re-writing, the analysis now follows that of the $\xi\to\infty$ and $\xi\to0$ limits made above: there is an exact parallel for each individual two-point function $$\langle\Psi_{p}|\boldsymbol\tau_{|\theta|,p}^{\alpha}(0,0)\bar{\boldsymbol\tau}_{|\theta|,p}^{\alpha}(x,t)|\Psi_{p}\rangle\quad (p=2q-1),$$ with the only difference that {\em it is not necessary to take the asymptotic limit in $\xi$}. For each $\theta$ (and each $p$), the factor $|\Psi_{|\theta|,p}\rangle$ of the full state $|\Psi_p\rangle$, on which $\tau_{|\theta|,q}^\alpha$ act non-trivially, correlates points $(x,t)$, $(x',t')$ only for
$$\frac{|x-x'|}{|t+t'|} \in [v(\theta-\epsilon/2),v(\theta+\epsilon/2)]$$ (recall that $\theta\in(\mathbb N+\frac12)\epsilon$). The analysis of single-mode correlations is made in App.~\ref{app:singlemode}.

Therefore, for $\xi>2 v(\theta+\epsilon/2)$ correlations occur on the horizontal path $(0,t)\to(x,t)$, but no correlations occur on $(0,t)\to(0,0)\to(x,0)\to(x,t)$ (note that, again, the segment of path $(0,0)\to(x,0)$ does not contribute). Thus we must choose the latter path (Fig.~\ref{quench_finitesystem} (left)). On the contrary, $\xi<2 v(\theta-\epsilon/2)$, correlation occur between the segment of paths $(0,t)\to(0,0)$ and $(x,0)\to(x,t)$, but not on the horizontal path $(0,t)\to(x,t)$. Thus we must choose the latter (Fig.~\ref{quench_finitesystem} (right)). In making these right choices, the correlation functions of pair-mode twist fields tend to their values in the long-time GGE,
\begin{eqnarray}
\label{eq:toGGEsingle}
    \lefteqn{\langle\Psi_{p}|\boldsymbol\tau_{|\theta|
    ,p}^{\alpha}(0,0)
    \bar{\boldsymbol\tau}_{|\theta|,p}^{\alpha}(x,t)|\Psi_{p}\rangle}&& \\ &\asymp&
    \begin{cases}
\langle \boldsymbol\tau_{|\theta|,p}^{\alpha}(0,t)\bar {\boldsymbol\tau}_{|\theta|,p}^{\alpha}(0,0)
\boldsymbol\tau_{|\theta|,p}^{\alpha}(x,0)\bar {\boldsymbol\tau}_{|\theta|,p}^{\alpha}(x,t)\rangle_{\rho_w}& \xi>2v(\theta+\epsilon/2)\\
\langle \boldsymbol\tau_{|\theta|,p}^{\alpha}(0,0)\bar{\boldsymbol\tau}_{|\theta|,p}^{\alpha}(x,0)\rangle_{\rho_w} & \xi<2v(\theta-\epsilon/2)\ .
\end{cases}\nonumber
\end{eqnarray}
Note how in the first line, it is a four-point function that appears.

Now again, in order to apply the BFT, we need to consider the correlations between twist fields within GGE. We already argued in Sec. \ref{ssect:renyihalf} (with supporting calculations in App.~\ref{app:approachGGE}) that no strong correlation occurs between local operators at equal times and different points in space, thus on the second line of \eqref{eq:toGGEsingle} we may apply the BFT. We also argued that no strong correlation occurs between current operators at equal space and different times, and in fact this also holds for single mode currents. However, in order to simplify the first line of \eqref{eq:toGGEsingle}, we need to address correlations between currents on the path segments $(0,t)\to(0,0)$ and $(x,0)\to(x,t)$. In general, local observables have strong correlations at different space-time points, due to hydrodynamic modes propagating in space-time. As we do not make any strong assumption about the dispersion relation, all hydrodynamic velocities occur, hence correlations occur between generic local observables on these separate path segments. 
However, single-mode currents only produce hydrodynamic modes at the corresponding velocities; supporting calculations are found in App.~\ref{ssect:decay}. Thus, as long as $\xi>v(\theta+\epsilon/2)$, no correlation occurs between these paths for the single mode currents $j_{\pm\theta,p}$. 
As $v(\theta)>0$, then $\xi>2v(\theta+\epsilon/2)\Rightarrow \xi>v(\theta+\epsilon/2)$, hence on the first line simplifies and we have
\begin{equation}\label{eq:toGGEsingle2}
    \langle\Psi_{p}|\boldsymbol\tau_{|\theta|
    ,p}^{\alpha}(0,0)
    \bar{\boldsymbol\tau}_{|\theta|,p}^{\alpha}(x,t)|\Psi_{p}\rangle \asymp
    \begin{cases}
\left| \langle \boldsymbol\tau_{|\theta|,p}^{\alpha}(0,t)\bar {\boldsymbol\tau}_{|\theta|,p}^{\alpha}(0,0)\rangle_{\rho_w} \right|^2 & \xi>2v(\theta+\epsilon/2)\\
\langle \boldsymbol\tau_{|\theta|,p}^{\alpha}(0,0)\bar{\boldsymbol\tau}_{|\theta|,p}^{\alpha}(x,0)\rangle_{\rho_w} & \xi<2v(\theta-\epsilon/2)
\end{cases}
\end{equation}
where the BFT can be applied for all two-point functions.

For the two-point functions with spatial separation $\langle \boldsymbol\tau_{|\theta'|,p}^{\alpha}(0,0)\bar{\boldsymbol\tau}_{|\theta'|,p}^{\alpha}(x,0)\rangle_{\rho_w}$, we can use the analysis of \eqref{eq:tautauspace} made in Sec.~\ref{subsec:spacedirection}, where we only have to replace, on the right-hand side of \eqref{eq:tautauspace} inside the $\theta$-integral, the constant one-particle eigenvalue $h_p$ by the piece-wise constant function $$h_p\;\big( \Theta(\epsilon/2 - |\theta-\theta'|) + \Theta(\epsilon/2 - |\theta+\theta'|)\big).$$ Thus the same analysis goes through, but with the integral restricted to $$\theta\in I_{\theta',\epsilon}:=[\theta'-\epsilon/2,\theta'+\epsilon/2]\cup [-\theta'-\epsilon/2,-\theta'+\epsilon/2].$$ Likewise for the two-point functions with temporal separation $\langle \boldsymbol\tau_{|\theta'|,p}^{\alpha}(0,t)\bar {\boldsymbol\tau}_{|\theta'|,p}^{\alpha}(0,0)\rangle_{\rho_w}$, with the analysis of Sec.~\ref{ssect:renyihalf}. Putting the results together, we obtain
\begin{equation}\begin{aligned}
    \frac1{1-\alpha}\log \Big(&\prod_{q=-\alpha/2+1}^{\alpha/2}\langle\Psi_{2q-1}|\boldsymbol\tau_{|\theta|,2q-1}^{\alpha}(0,t)\bar{\boldsymbol\tau}_{|\theta|,2q-1}^{\alpha}(x,t)|\Psi_{2q-1}\rangle\Big)\\
    &\sim
    \int_{I_{\theta,\epsilon}}\frac{d\theta'}{2\pi}\min(x,2t|v(\theta')|)\:H_{\alpha}(\theta')
    \end{aligned}
\end{equation}
which is valid for $x/t <2v(\theta-\epsilon/2)$ or $x/t >2v(\theta+\epsilon/2)$.

For the case of $x/t$ within this excluded region, we do not have explicit results, but the scaled cumulants still are finite (as one can see by doing a calculation similar to  App.~\ref{app:approachGGE}, for instance). Thus, the result may be deemed valid as well within this excluded region, up to an error of order $\epsilon$.

Taking the product over $\theta$'s as per \eqref{eq:U1_twist_psi_single},
\begin{equation}
    \frac1{1-\alpha}\log\langle\Psi^\alpha| T^{\alpha}(0,t)\bar{T}^{\alpha}(x,t)|\Psi^\alpha\rangle
    \asymp
    \int\frac{d\theta}{2\pi}\min(x,2t|v(\theta)|)\:H_{\alpha}(\theta)
    + O(\epsilon)
\end{equation}
and as this holds for all $\epsilon>0$, we can take the limit $\epsilon\to0$ and we obtain
\begin{equation}
S_{\alpha}(x,t)=\frac{1}{1-\alpha}\log\langle T^{\alpha}(0,t)\bar{T}^{\alpha}(x,t)\rangle \sim\int\frac{d\theta}{2\pi}\min(x,2t|v(\theta)|)\:H_{\alpha}(\theta)\ .\label{eq:fullRenyi}
\end{equation}
This is in full agreement with the quasiparticle picture \cite{alba2017entanglement,alba2018entanglementlong}.

Finally, we note that the relation \eqref{eq:mainfluctu} between this formula for R\'enyi entanglement entropy growth, and the static and dynamic fluctuations, is directly obtained from the above discussion, by identifying
\begin{equation}
    J_{N_<}(t) = \int_0^t dt'\,\sum_{\theta\in(\mathbb N+\frac12)\epsilon\atop 2v(\theta)<x/t} j_{|\theta|}(0,t')
\end{equation}
and
\begin{equation}
    N_>(x) = \int_0^x dx'\,\sum_{\theta\in(\mathbb N+\frac12)\epsilon\atop 2v(\theta)>x/t} q_{|\theta|}(x',0)
\end{equation}
using the explicit expressions of pair-mode densities and currents \eqref{eq:pairmodedensities} in App.~\ref{app:singlemode}, and taking the limit $\epsilon\to 0$.

\section{Discussion and conclusion}
\label{sec:discussion}

In this paper we have studied the R\'enyi entanglement entropy in GGEs and after quenches from integrable (pair-production) states in free fermion theories. Although this has been relatively well studied in the literature, most results were based on specific ways of writing the R\'enyi entanglement entropy using the free fermion structure (e.g.~in terms of determinants), and on the idea of entanglement due to engangled pairs produced by the quench \cite{alba2017entanglement,alba2018entanglementlong}. A first-principle derivation that generalises beyond free fermions was still largely missing, while it is known that the simple quasi-particle picture fails for $\alpha$-R\'enyi entanglement entropies (with $\alpha\neq 1$) in interacting models \cite{bertini2022growth}.

We have proposed a new approach based on twist-field correlation functions and hydrodynamic fluctuations. This uses the hydrodynamic theory for free fermions, which is a special case of generalised hydrodynamics (GHD), and the ballistic fluctuation theory (BFT), which relates the exponential decay of twist-field correlation functions to hydrodynamic large-deviation theory. Crucially, in order to have a full understanding of the quench dynamics, we have introduced a new concept: that of single-mode twist fields. These are twist fields associated to the quasi-local charge counting the number of fermions within a small interval of momentum; or more generally twist fields ``acting" on the quasi-locality sector of observables supported on a small momentum interval. The approach is potentially more general and more fundamental, as hydrodynamics, the BFT and single-mode twist fields -- twist fields associated to individual hydrodynamic modes -- are applicable much beyond free fermions. Perhaps most interestingly, it reveals the new physics of thermodynamic and hydrodynamic fluctuations behind the behaviour of the R\'enyi entanglement entropies.

Three important concepts are brought forward:
\begin{itemize}
    \item Entanglement is deeply connected to fluctuations, and the large-scale behaviour of entanglement, both static and dynamic, is controlled by large-deviation and hydrodynamic principles.
    \item Hydrodynamic modes and projections onto such modes are more accurate and general notions which replace the idea of particle-pair productions used to understand entanglement dynamics in integrable systems.
    \item The fact that the entanglement growth in quenches from ``integrable'' states can be written as a simple and universal function of the generalised Gibbs ensemble (GGE) reached at long times comes from the particularly simple structure of the {\em long-range correlations} that such states present.
\end{itemize}

To elaborate on these concepts: first we have confirmed that the R\'enyi entanglement entropy in GGEs is controlled by thermodynamic fluctuations, and related to (a simple analytic continuation of) a difference of thermodynamic free energies. This is in agreement with the observations, made earlier \cite{PhysRevLett.102.100502,PhysRevB.85.035409,Calabrese_2012}, that the large-deviation theory for charge fluctuations is closely related to the entanglement entropy. 
Here, this relation appears naturally from completely general concepts: branch-point twist fields and the BFT. Using these, in fact, the conclusion is pushed further: we show that the growth of R\'enyi entanglement entropy after a quench is controlled by hydrodynamic current fluctuations, and related to a dynamical free energy associated to the large-deviation theory for charge transport, as fully encoded in Eq.~\eqref{eq:mainfluctu} in the introduction (we mention that some qualitative arguments in this direction were already present in \cite{jin2021interplay}). 
The relation between R\'enyi entanglement entropy and charge fluctuations is a general aspect of quadratic theories (not only free fermions, as our approach could also be generalized to free bosons), as in such theories, the branch-point twist field can be written as a product of $U(1)$ twist fields, which are then associated, by the BFT, to the large-deviation theory of $U(1)$ charge fluctuations. 

Second, the methods we have developed show that the notion of quasi-particle used in integrable systems to explain the behaviour of entanglement, should in fact be replaced by that of hydrodynamic mode. Indeed, the BFT, which describes the exponential behaviour of twist field correlation functions, is purely based on the Euler hydrodynamic data of the microscopic model. In free fermion models, and in integrable models, it turns out that hydrodynamic modes are in one-to-one correspondence with quasi-particles (see e.g.~the review\cite{DeNardis_2022}); and in particular, the single-mode twist fields we have introduced, are twist fields associated with such hydrodynamic modes. But beyond these situations, hydrodynamic modes are the more general objects at play in the large-scale dynamics of many-body systems.

Third, our calculations explain why it is possible to specify the growth and saturation of entanglement after quenches from pair-production states in a simple way in terms of the long-time GGE. This is not based on the conventional physical picture of entanglement produced by entangled pairs of particles.  But rather, it is based on the study of {\em long-range correlations that such pairs give rise to after quenches}. It has not been appreciated until now that quenches give rise to long-range spatial correlations 
, of the type found recently in non-equilibrium, long-wavelength states \cite{doyon2022ballistic,doyon2022emergence}. These long-range correlations are generically seen by observables supported on regions of space that are large enough; specifically, with a ballistic scaling of the region's length $x$ with respect to the time $t$ since the quench. Thus, for such observables, the state is not a GGE. This is important, as twist fields are semi-local with respect to the fermions, thus the semi-locality branch is affected by such correlations.

The fact that the long-time GGE can be used to describe not only the saturation of the R\'enyi entanglement entropies but also their growth in a simple way, is {\em because of the particular structure of long-range correlations in integrable pair-production quenches}.
Indeed, in such quenches, for every choice of $x/t$, one can always choose a path in space-time which avoids all long-range correlations. This follows from a simple geometric analysis of trajectories in space-time. Intuitively, long-range correlations occur between positions in space-time where pairs of correlated particles lie, a picture that we fully support by simple calculations of correlation functions of conserved densities and currents and their asymptotic behaviours. Once the path avoids long-range correlation, it only perceives the long-time GGE.

Note that it is obvious that the entanglement growth can be described purely in terms of the final GGE (with no further information from the initial state needed), because of the one-to-one correspondence between initial squeezed state (pair-production state) and GGEs, see Eq.~\eqref{eq:quenchgge} (so no more information about the initial state is present at all). The structure of long-range correlations however allow a simple description, in terms of fluctuations within the long-time GGE. In general, the one-to-one correspondence is lost in quenches from more complicated states, and the universality of the entanglement growth is also lost \cite{bertini2018entanglement,Bastianello_2018}.

Importantly, we find that the branch-point twist field can be decomposed into {\em tensor factors -- the single-mode twist fields -- that act on each small momentum interval}. Each small momentum interval is associated with a family of quasi-local operators, with respect to which branch-point twist fields can be defined\footnote{In our calculation, we used the explicit free-fermion basis to first write the branch-point twist fields in terms of $U(1)$ twist fields by the standard arguments of \cite{cardy2008form}, which we then factorised into single-mode twist fields. But the concept remains valid without the mapping to $U(1)$ twist fields.}. Each such twist field is only semi-local with respect to the fermions pertaining to the single momentum interval, and, pairing intervals of opposite momenta, allowed us to separate the two-point function of branch-point twist field (used to evaluate the R\'enyi entanglement entropy) into a product of contributions on different momentum intervals. For each factor, the semi-locality branch can be chosen in order to avoid long-range correlations corresponding to the pair it perceives. We provided extensive calculations of correlation functions of single-mode densities and currents that support this physical picture.

From these considerations we fully reproduced the long space-time dynamics of the R\'enyi entanglement entropy that had been obtained by pair-entanglement argument.

Many extensions of our work are possible. Most importantly, we believe the derivation we have provided, and many of the conclusions, can be extended to interacting integrable models, and potentially to interacting non-integrable models.

In interacting integrable models, it would be interesting to reproduce, and provide a better understanding of, the recent result \cite{bertini2022growth}. This was obtained using crossing symmetry of relativistic quantum field theory, in order to relate R\'enyi entanglement entropy growth in time to the linear scaling of R\'enyi entanglement entropy in space, much like one can evaluate currents by crossing from conserved densities \cite{castro2016emergent}. However, as we have argued, the more general understanding of time-extensive behaviours is via hydrodynamic modes. Thus, it is likely that hydrodynamic ideas will provide a more first-principle derivation.

Technically, in interacting models, it is not possible to factorise the branch-point twist fields, in such a simple way as in \cite{cardy2008form}, into $U(1)$ twist fields, a trick that we have used here. We believe this difficulty can be surmounted as follows. First, it is still possible to diagonalise the twist field action, at the price of making the resulting S-matrix non-diagonal, see App.~\ref{app:Salpha_matrix}. The branch-point twist field is then associated with a symmetry that has diagonal action on the new asymptotic particles, and hydrodynamic modes can be constructed from these particles (by constructing the corresponding nested thermodynamic Bethe ansatz). Having twist fields associated to charges that are diagonal in the particle basis, the results of the BFT for generic intergable models can in principle be applied.

It is also immediate that single-mode twist fields exist as well in interacting integrable models, which would be interesting to study, independently form their applications to entanglement.

Another avenue is to use the ballistic macroscopic fluctuation (BMFT) theory developed recently \cite{doyon2022ballistic}. This is a more general construction which does not involve a flow on GGEs (by constrast to the BFT). This would allow us to apply the principles introduced here -- the relation between hydrodynamic fluctuations and entanglement entropy using twist fields -- beyond homogeneous quantum quenches, and beyond the simple particle-pair quenches, as the BMFT is applicable to inhomogeneous situations and for generic long-rance correlation structures. In particular, it would be interesting to account for the long-range correlations not by choosing paths that avoid them, but by evaluating directly their influence on the fluctuations. This would be important, as initial states that are inhomogeneous generically produce long-range correlations \cite{doyon2022ballistic,doyon2022emergence} that are not necessarily of particle-pair type. This could also access quenches where multiple-particle processes are involved \cite{bertini2018entanglement, Bastianello_2018}.

Beyond integrability, perhaps the main results are those based on a notion of ``surface tension" underlying entanglement entropy in chaotic systems \cite{PhysRevX.10.031066}. It would be interesting to apply our methods, based on hydrodynamics, to such situations.

A natural further extension is to introduce the effects of diffusion, in particular using the exact results in integrable models \cite{PhysRevLett.121.160603}, and potentially the effects of dispersion \cite{denardishigher}.

Finally, there is no reason to restrict ourselves to entanglement entropy, or to quantum systems. The methods we have introduced here are immediately applicable, for instance, to the large-deviation theory of $U(1)$ densities and currents after quenches in interacting models, both at and away from integrability, and both for quantum and classical systems. This would be a very interesting direction to investigate.




\section*{Acknowledgements}
We thank Vincenzo Alba and Olalla Castro-Alvaredo for discussions and collaborations on closely related subject.

\paragraph{Funding information}
The work of BD was supported by the Engineering and Physical Sciences Research Council (EPSRC) under grants EP/W000458/1 and EP/W010194/1.


\begin{appendix}

\section{Remarks on notions of locality and twist fields}\label{app:remarks}

There is a lot more that one can say about the twist fields we introduced in Sec.~\ref{subsec:twistfields}, as well as the notions of locality we briefly discussed. Here we collect a number of remarks in order to provide a brief and rough guide to this wide subject.

\subsection{Unbounded observables and topological charges}

As is made clear from the height field definition of twist fields, a twist field may exist as soon as there is an observable, say $\varphi(x,t)$, that is ``unbounded": that takes values in a non-compact space which are not bounded by the dynamics. When this happens, the field can grow from a value it has at $x,t$ to another well separated value at $x',t'$. If this growth is linear, then from this we can construct a twist field as done in Sec.~\ref{subsec:twistfields}, with a large-deviation theory described by the BFT. This usually happens when there is a non-compact symmetry\footnote{This is essentially a ``gauge symmetry": non-compactness is the main aspect of gauge invariance that makes is different from oridnary symmetries.}, such as the $\mathbb Z$ symmetry group of the sine-Gordon model if the sine-Gordon field is taken in $\mathbb R$, or the $\mathbb R$ symmetry group of the real free massless Boson. In such cases, $\varphi(\infty)-\varphi(-\infty)$ is a ``topological charge", it is an extensive conserved quantity with density $q(x,t) = \partial_x \varphi(x,t)$. The vertex operators $e^{-\ii \eta\varphi(x,t)}$ are the associated twist fields, and the extensive charge $\varphi(\infty)-\varphi(-\infty)$ should be considered as part of the space of extensive charges $Q_i$ used to construct GGEs. It will appear after appropriate quenches via (generalised) thermalisation.

\subsection{Descendant twist fields and semilocality sectors} The exchange relations \eqref{exchangerelationgeneral} are a good way of characterising twist fields. However, they characterise not a single field, but a {\em family of fields}. Indeed, clearly, the identification \eqref{eq:identificationtwist} is not the unique choice satisfying \eqref{exchangerelationgeneral}. For instance, $$T(x,t) = a(x,t) T_{-\ii \eta}(x,t)$$ will also work, for any local observable $a(x,t)$. The choice \eqref{eq:identificationtwist} may be seen as a ``highest-weight" twist field, and the above are usually referred to as ``descendant twist fields" (these notions make full sense, for instance, in quantum or conformal field theory, using the concept of dimension). All such descendants are in the same ``semilocality sector" $\mathcal T$ defined by the exchange relation \eqref{exchangerelationgeneral}. One application of the BFT to descendant twist fields is explained in \cite{del2022hydrodynamic} in the context of the XX quantum chain.

\subsection{Non-abelian semilocality} Given two ``local enough" symmetry transformations $\sigma$ and $\sigma'$, that is $a(x,t)\mapsto \sigma a(x,t)$ and $a(x,t)\mapsto \sigma'a(x,t)$, we can define two semilocality sectors  $\mathcal T_\sigma$ and $\mathcal T_{\sigma'}$. In general, if the transformations do not commute, one has that if $T'\in \mathcal T_{\sigma'}$, then $\sigma T' \in \mathcal T_{\sigma\circ\sigma'\circ\sigma^{-1}}$. Further, if $T\in \mathcal T_{\sigma}$ and $T'\in \mathcal T_{\sigma'}$, then the exchange relation takes the form
\begin{equation}\label{exchangerelationmostgeneral}
    T'(x,t) T(y,t) = \left\{\begin{array}{ll}
         T(y,t) \,(\sigma^{-1}T')(x,t) & (y\ll x) \\
        (\sigma' T)(y,t)\, T'(x,t)& (y\gg x).
    \end{array}\right.
\end{equation}

\subsection{Twist fields in the literature} 

It is difficult to give a full account of the literature on twist fields. As a guidance we mention that twist fields and their semilocality have been discussed extensively in various contexts, including: phase transitions in classical and quantum statistical models \cite{PhysRevB.3.3918,PhysRevD.15.2875,SCHROER197880} (see the review \cite{fradkin2017disorder}); vertex operators, Yangians, parafermions and orbifolds in conformal and integrable quantum field theory \cite{cmp/1104178892,SmirnovBook,bernardCMP,BERNARD1993709,BERNARD1994534}; tau-functions and Painlev\'e equations \cite{holonomicI,holonomicII,holonomicIII,holonomicIV,holonomicV,Palmer,DOYON2003607,lisovyy}; and entanglement entropy in quantum field theory and in quantum spin chains \cite{cardy2008form,castro2011permutation,PhysRevLett.109.130502,PhysRevLett.120.200602}. Twist fields have also been considered in higher dimensions \cite{hungtwist}. In most works, the focus is on ultra-local ``internal" symmetries, that strictly factorise in space, usually part of a symmetry group such as $\mathbb Z_n$, $U(1)$, $SU(n)$, permutations, etc. Note that for ultra-local symmetries, large inequalities $\ll,\,\gg $ can be replaced by ordinary inequalities $<,\,>$ in \eqref{exchangerelationgeneral} for finite distances between the supports of the observables involved. This is the usual way of writing the exchange relations. More recently, twist fields associated to spacetime boost transformations in QFT, that are not ultra-local, have been considered \cite{CASTROALVAREDO2018146}; this is an example where the hamiltonian density is not preserved by the transformation, $\tilde h(x,t) \neq h(x,t)$.

\subsection{Concepts of locality in the literature} The concept of ``locality" has been discussed widely in the literature, under a variety of definitions. In relativistic QFT, local fields are those that commute at space-like distances with the energy-momentum tensor. It is important to remark that under this general definition, local fields include twist fields associated to internal symmetries. This definition can in fact be used in any quantum model, be it a field theory, spin chain or model of interacting particles. In fact, one defines ``locality sectors" containing families of local fields that commute with each other at space-like distances; and a distinguished locality sector is that containing the energy density. These considerations are at the basis of orbifold conformal field theory \cite{cmp/1104178892}.

In spin chains, the most na\"ive concept is that of operators supported on finitely many sites. In the $C^*$ algebra formulation, this is completed with respect to the operator norm \cite{bratteli2012operator}; importantly, the property of operators commuting in the limit of large separations is preserved by this completion. Part of these $C^*$ algebra elements are the ``quasi-local operators" that have been introduced in order to describe generalised thermalisation in integrable models \cite{ilievski2016quasilocal}. These are elements of the $C^*$ algebra for which one can still define a finite support, but only up to corrections of exponentially decaying norm.

But much like in QFT, twist fields, which are semilocal with respect to generic observables but may be local with respect to some family of observables including the energy density, can also be adjoined to the $C^*$ algebra, as is natural to do for instance in the context of Jordan-Wigner transformations \cite{arakiXY}. Then, either from the $C^*$ algebra, or from some potentially smaller algebra of observables deemed local (for instance with appropriate decay of correlation functions, and which, again, may include twist fields), other completions are possible, and sometimes more physically relevant. For instance, the Gelfand-Naimark-Segal Hilbert space with respect to a given state, and its space of bounded operators, both are usually larger than the $C^*$ algebra. Another Hilbert space completion is that based on susceptibilities \cite{doyonpseudo}. This gives the concept of ``extensive" quantities, a generalisation of quantities written as sums over space of local operators (or ``local densities"). These form a Hilbert space, {\em a priori} without any algebraic structure. In particular, it has been shown \cite{doyonpseudo} that extensive conserved quantities are in bijection  with ``pseudolocal charges", roughly defined by their extensivity property $\langle Q^2\rangle^c \propto L$ in a system of length $L$ \cite{ilievski2016quasilocal}. Extensive conserved charges form a complete set which has been rigorously shown to fully describe the linearised Euler hydrodynamics \cite{doyonlinearised}, and they may be used to more formally and precisely construct GGEs (addressing convergence issues) \cite{doyonpseudo}.

Despite all these studies, the full relation between extensive conserved quantities, twist fields, local symmetry transformations and GGEs is still not fully unexplored.

\section{Correlations after a quench from an initial state with pair structure} \label{app:correlations}

In this appendix we provide supporting argument for the choice of the integration path for evaluating the SCGF made in the main text. We recall that the choice of path is dictated by two main ideas:
\begin{itemize}
    \item The production of pairs of quasi-particles by the initial state by the after-quench dynamics gives rise to long-range correlations: points reached by a pair of quasi-particles with opposite momenta are correlated. These quasi-particles have the interpretation of fluid modes, and such long-range hydrodynamic correlations are akin to those seen in the Ballistic Macroscopic Fluctuation Theory (BMFT) \cite{doyon2022ballistic,doyon2022emergence}. 
    \item BFT \cite{doyon2020fluctuations} is applicable only when multi-point correlation functions of the densities and currents, the integrand in \eqref{eq:Delta_j}, cluster fast enough along the chosen contour. Otherwise, depending on the structure of such correlations, the SCGF and the cumulants may be divergent under ballistic scaling, or the BFT result may receive additional contributions from these correlations which have to be taken into account. The more general BMFT \cite{doyon2022ballistic} in principle provides the corrections. However, it is simpler to directly apply the BFT by choosing contours that avoid such correlations, leading to the paths shown in Fig. \ref{quench_finitesystem}.
\end{itemize}

Below, we explicitly show the presence/absence of such long-range correlations along the different paths considered in the main text.

We use the notations $a(x,t)$ and $b(x,t)$ for free fermionic fields with the usual normalisation, e.g.
\begin{equation}\label{bnormalisation}
    \{b^\dagger(x),b(y)\} = \delta(x-y),\quad
    \{b^\dagger_\theta,b_\alpha\} = \delta(\theta-\alpha),\quad
    b(x,t) = \int \frac{d \theta}{\sqrt{2\pi}}
    e^{\ii x \theta} b_\theta(t)
\end{equation}
where $b_\theta(t) = e^{-\ii E(\theta)t}b_\theta$. The global $U(1)$ charge is
\begin{equation} \label{Qappendix}
    Q = \int d\theta\,b_\theta^\dagger b_\theta.
\end{equation}
Note that in the main text a variety of canonical free fermion fields were defined: the original fields $\psi(x,t)$, the replicated ones parametrised by a copy index $\psi_i(x,t)$, and the fields obtained from these by diagonalising in copy space, $\psi_p(x,t)$. These all are canonical free fermion fields (independent from each other for different copy number $i$, or for different diagonalised copy number $p$). The calculations below therefore apply to any such choice of fields.

As we discuss quenches, in this appendix we use two consecutive letters for the canonical free fermion fields: $a(x,t)$ (for the pre-quench fermion) and $b(x,t)$ (for the post-quench fermion).

\subsection{Global $U(1)$ densities and currents and decay of correlations in GGEs} \label{app:global}

First, recall the definition of the generalised current as a line integral \eqref{eq:Delta_j}
\begin{equation}
    \Delta J(\gamma) = \int_{\underline \ell} (j(x,t)dt - q(x,t)dx)\label{eq:appendixA:scgf}
\end{equation}
where $j(x,t)$ and $q(x,t)$ are the current and the density associated the $U(1)$ conserved charge $Q$ in \eqref{Qappendix}. We recall also that the above integral only depends upon the end points of the path $\underline \ell$ due to the conservation law relating current and density, $\partial_t q(x,t) + \partial_x j(x,t) = 0$. In free models with global $U(1)$ symmetry the fermionic Hamiltonian is of the form
\begin{equation}
    H=\int d\theta \,E(\theta) b^\dag_\theta b_\theta \label{eq:appendixA:hamiltonian}
\end{equation}
where $E(\theta)$ is the dispersion relation (recall that $E(\theta) = E(-\theta)$ is a strictly convex function). The local charge density is given by
\begin{equation}
    q(x,t) = b^\dag(x,t) b(x,t)\label{eq:appendixA:local_density}
\end{equation}
and it can be easily verified by using the fermionic algebra that $Q=\int dx\, q(x,t)$ is conserved, $\commutator{Q}{H}=0$. Let us find the associated local current.

By the conservation law we have (we suppress the time dependence as all fields are the same time $t$)
\begin{align}
    \partial_x j(x) &= - \partial_t q(x) = \ii \commutator{q_x}{H} = \ii\int d\theta E(\theta) \commutator{b^\dag(x) b(x)}{b^\dag_\theta b_\theta}
    \nonumber
    \\
    & = \ii\int d\theta \frac{dk}{\sqrt{2\pi}}\frac{dk'}{\sqrt{2\pi}} E(\theta) e^{-\ii x(k - k')}\commutator{b^\dag_kb_{k'}}{b^\dag_\theta b_\theta}
    \nonumber
    \\
    &= \ii\int d\theta  \frac{dk}{2\pi}E(\theta) \left(e^{-\ii x(k-\theta)}b^\dag_k b_\theta - e^{-\ii x(\theta-k)}b^\dag_\theta b_k\right)
    \nonumber
    \\
    &=\ii\int d\theta  \frac{dk}{2\pi} e^{\ii x(k-\theta)}\left(E(k) - E(\theta)\right)b^\dag_\theta b_k  .
\end{align}
Integrating with respect to $x$ we find
\begin{equation}
    j(x,t) = \frac{1}{2\pi}\int d\theta dk\, e^{\ii x(k-\theta)}\left(\frac{E(k)-E(\theta)}{k-\theta}\right)b^\dag_\theta(t) b_k(t)\label{eq:appendixA:local_current}
\end{equation}
where the $x$-independent integration constant is chosen in such a way that the result is a local observable\footnote{This in fact fixes the result up to an overall term proportional to the identity operator $\mathbf 1$; indeed there are no $x$-independent homogeneous local operators, whose space-time translations are generated by the momentum and Hamiltonian, other than  $\bf 1$.}. This is the known expression for the current in the case of a quadratic dispersion relation in the continuum. Actually, restricting the integration over momenta in $[-\pi,\pi]$ and taking $E(k)=\cos(k)$ one reproduces also the current on the lattice; but here we keep $\theta,k\in\mathbb R$ for simplicity.

We now show that in a GGE, the connected correlation functions of densities decay fast enough in space, and the correlation functions of currents decay fast enough in time, in such a way that scaled cumulants are finite, thus making the BFT applicable. The former in fact is valid for all local observables, while the latter only hold for the currents.

For simplicity, here and in the following subsections, we will concentrate solely on two-point correlation functions -- although all higher-point functions (and their respective cumulants) should in principle be investigated similarly.

Let $\braket{\cdot}$ be a GGE. Let us assume that the occupation function $n(\theta)$ characterising the GGE is analytic in a neighbourhod of $\mathbb R$. Using
\begin{equation}
    \braket{b^\dagger_\theta b_{\theta'}}
    =\delta(\theta-\theta') n(\theta)
    \label{eq:bdagb_GGE}
\end{equation}
we have, on the one hand,
\begin{equation}
    \braket{b^\dagger(x)b(0)}
    = \int \frac{d\theta}{2\pi}\,e^{-\ii x\theta}n(\theta)\quad .
\end{equation}
For $x>0$ (resp.~$x<0$), contour deformation can be performed as $\theta\mapsto \theta - \ii \gamma$ (resp.~$\theta\mapsto \theta + \ii \gamma$) for $\gamma>0$ small enough, and we see that the resulting integral decays exponentially as $|x|\to\infty$. This implies exponential decay of all two-point connected correlation functions of local observables formed out of sums of products of $b(x),\,b^\dagger(x)$ and their derivatives, including $U(1)$ densities. It also implies linear scaling of cumulants; for instance this would mean 
\begin{align}
    &\int_0^X dx \int_0^X dx' \braket{b^\dag(x) b(x')} 
    \nonumber
    \\
    &=\int_0^X dx \int_0^X dx' \int \frac{d\theta}{2\pi}\,e^{-\ii (x-x')\theta}n(\theta)
    \nonumber
    \\
    &\sim \int_0^X dx \int_0^X dx' e^{-\gamma |x-x'|}\sim X\label{eq:cumulant_scaling}
\end{align}
where in the last line we have shifted $\theta\mapsto \theta - \ii \sign(x-x')\gamma$ and used $\sign(x) x = |x|$.
This is the correct ballistic growth of the cumulant.

On the other hand, we find
\begin{equation}
    \braket{b^\dagger(0,t)b(0,0)}
    = \int \frac{d\theta}{2\pi}\,e^{\ii tE(\theta)}n(\theta).
\end{equation}
This has a stationary phase at $\theta_*$ such that $E'(\theta_*)=0$; this point is unique by our assumption of strict convexity (and $\theta_*=0$ by symmetry, although we don't make use of this fact in this calculation), so a saddle point analysis gives
\begin{equation}
    \braket{b^\dagger(0,t)b(0,0)}
    \sim \frac{\sqrt{\ii}\,e^{\ii t E(\theta_*)} \,n(\theta_*)}{\sqrt {2\pi t}}.
\end{equation}
Therefore, correlation functions of generic local observables $o(x,t),\,o'(x,t)$ formed out of {\em bilinears of creation and annihilation operators} have algebraic decay
\begin{equation}
    \braket{o(0,t)o'(0,0)}^c = O\big(\frac1{t}\Big)\qquad (t\to\infty).
\end{equation}
For such decay, cumulants of total time integrals do not grow linearly, 
\begin{equation}
    \braket{\int_0^T dt\,o(0,t)\,\int_0^T dt'\,o'(0,t')}^c \gg T\qquad(T\to\infty)
\end{equation}
thus breaking the large-deviation principle at the basis of the BFT. However, an important remark is that this generic behaviour of fermion bilinears {\em does not hold in the case of currents, $o(x,t)=o'(x,t)=j(x,t)$}. Indeed, using \eqref{eq:appendixA:local_current} with $x=0$, we see that we must set $\theta=k=\theta_*$ for the long-time limit of the current two-point function. From
\begin{equation}\label{eq:EEk}
    \frac{E(k)-E(\theta)}{k-\theta}
    = E'(k) + O(k-\theta)
\end{equation}
we realise that $\frac{E(k)-E(\theta)}{k-\theta}\big|_{k=\theta=\theta_*} = 0$. Therefore, the current two-point function decays faster than $1/t$; in fact it decays as
\begin{equation}\label{eq:decay_current_current_infinite_interval}
    \braket{j(0,t)j(0,0)}^c = O\big(\frac1{t^3}\Big)\qquad (t\to\infty).
\end{equation}
This guarantees the correct scaling of cumulants
\begin{equation}\label{eq:jjT}
    \langle \int_0^T dt\,j(0,t)\,\int_0^T dt'\,j(0,t')\rangle^c = O(T)\qquad(T\to\infty)
\end{equation}
and thus the validity of the BFT.

A similar argument shows that the current perpendicular to the path $\underline\ell$ -- the integrand in \eqref{eq:appendixA:scgf} -- has a similar property along the path, thus guaranteeing that the clustering requirement \eqref{clustering2} holds.

\subsection{Quench protocol and initial state} \label{ssect:quench}
In order to describe the quench protocol considered in the main text (for which we obtain predictions on the dynamics of the entanglement entropy) we define the pre-quench and the post-quench fermionic Hamiltonians respectively as
\begin{align}
    H_0 &= \int d\theta\, E_0(\theta) a^\dag_\theta a_\theta
    \\
    H &= \int d\theta\, E(\theta) b^\dag_\theta b_\theta
\end{align}
where again the fermions satisfy $\{a_{\theta}, a^\dag_{\theta'}\}=\delta(\theta-\theta')$, $\{b_{\theta}, b^\dag_{\theta'}\}=\delta(\theta-\theta')$.  According to the protocol, the system is initialized in the ground state of $H_0$ and then let evolve with $H$.
This corresponds to changing the whole dispersion relation (not only a parameter as in typical quenches in the literature), see e.g.~\cite{Essler_2016,doi:10.1146/annurev-conmatphys-031016-025451}.

The two set of fermions are related by a Bogolioubov-type transformation in the following way
\begin{equation}
    \begin{pmatrix}
    a_\theta
    \\
    a^\dag_{-\theta}
    \end{pmatrix}
    = 
    \begin{pmatrix}
    f_\theta & g_\theta\\
    g^*_{-\theta} & f^*_{-\theta}
    \end{pmatrix}
    \begin{pmatrix}
    b_\theta
    \\
    b^\dag_{-\theta}
    \end{pmatrix} .
    \label{appendixA:bogolioubov}
\end{equation}
Imposing the validity of anticommutation relations one gets the following constraints on the functions $f_\theta, g_\theta$
\begin{subequations}\label{eq:fgconstraints}
    \begin{equation}
    f_\theta g_{-\theta} + f_{-\theta} g_{\theta} = 0
    \end{equation}
    \begin{equation}
    |f_\theta|^2 + |g_{\theta}|^2= 1 .
    \end{equation}
\end{subequations}
Note that the first of these is identically satisfied choosing $f_\theta=f_{-\theta}$ and $g_\theta=-g_{-\theta}$ or viceversa. In our analysis we will keep these functions general.

The initial state $\ket{\Psi}$ is defined as
\begin{equation}
    a_\theta\ket{\Psi} = 0
\end{equation}
which, for instance, could be a filled Fermi sea so that operators $a_\theta$ are to be interpreted as creating excitations on top of these. It can be easily shown that in terms of post-quenches quantities this is described by the following squeezed state
\begin{equation}\label{squeezed}
    \ket{\Psi} = \frac1{\mathcal N}\exp\left(\frac12 \int d\theta\,\mathcal{K}_{\theta,-\theta}b^\dag_\theta b^\dag_{-\theta}\right)\ket{0}
\end{equation}
where $\ket{0}$ is the ground state of the post-quench Hamiltonian satisfying $b_\theta\ket{0} = 0$ (this state has the nice property of being \emph{gaussian} so that Wick's theorem applies). The function $\mathcal{K}_{\theta,\theta'} = -\mathcal{K}_{\theta',\theta}$ can be related directly to the functions $f_\theta$ and $g_\theta$ appearing in \eqref{appendixA:bogolioubov} using the fact that $\ket{\Psi}$ is annihilated by $a_\theta$. 
In terms of post-quench operators, this condition reads
\begin{equation}
    (f_\theta b_\theta + g_\theta b^\dag_\theta)\exp\left(\frac 12\int d\theta'\mathcal{K}_{\theta', -\theta'}b^\dag_{\theta'} b^\dag_{-\theta'}\right)\ket{0} = 0  .
    \label{eq:appendixA:relationKfg}
\end{equation}
 Using the BCH formula $e^A B e^{-A} = e^{[A,\circ]}B$ we obtain
 \begin{align}
     \exp\left(-\frac12\int d\theta'\mathcal{K}_{\theta', -\theta'}b^\dag_{\theta'} b^\dag_{-\theta'}\right)&b_\theta\exp\left(\frac12\int d\theta'\mathcal{K}_{\theta', -\theta'}b^\dag_{\theta'} b^\dag_{-\theta'}\right) 
     \nonumber
     \\
     &= b_\theta + \frac12\mathcal{K}_{\theta, -\theta}b^\dag_{-\theta} - \frac12\mathcal{K}_{-\theta, \theta}b^\dag_{-\theta}
     \nonumber
     \\
     &= b_\theta + \mathcal{K}_{\theta, -\theta}b^\dag_{-\theta} 
 \end{align}
which used in \eqref{eq:appendixA:relationKfg} in combination with $b_\theta\ket{0}=0$ gives
\begin{equation}
    [f_\theta(b_\theta + \mathcal{K}_{\theta, -\theta}b^\dag_{-\theta}) + g_\theta b^\dag_{-\theta}]\ket{0} = 0
\end{equation}
so that the condition is
\begin{equation}
    \mathcal{K}_{\theta, -\theta} = -\frac{g_\theta}{f_\theta} .
\end{equation}
Note in particular that we must have, by the anti-symmetry $\mathcal K_{\theta, -\theta} = - \mathcal K_{-\theta, \theta}$,
\begin{equation}\label{eq:g0}
    g_0 = 0.
\end{equation}

Finally, we may evaluate the predicted long-time GGE for the quench simply by evaluating the post-quench conserved quantities in the pre-quench vacuum state $|\Psi\rangle$. Inverting \eqref{appendixA:bogolioubov}, we write 
\begin{equation}\label{eq:babogo}
    b_\theta = \frac{f^*_{-\theta}a_\theta - g_\theta a^\dagger_{-\theta}}{f_\theta f^*_{-\theta} - g_\theta g^*_{-\theta}}.
\end{equation}
For this computation, it will be convenient to look at the values of the extensive conserved quantities;  hence we take a finite system of length $L$. This is warranted, as the quench is homogeneous. The above description of the quench stays valid with the discretisation $\theta\in\mathbb Z 2\pi/L$, and with the usual canonical anti-commutation relations with regularisation $\delta(\theta-\theta') \to \delta_{\theta,\theta'} \frac{L}{2\pi}$. We consider $b^\dagger_\theta b_\theta$, and obtain, after some algebra using Eqs.~\eqref{eq:fgconstraints},
\begin{equation}
    b^\dagger_\theta b_\theta =a^\dagger_{\theta}a_\theta |f_{-\theta}|^2 + a_{-\theta}a_{-\theta}^\dagger |g_\theta|^2\ .
\end{equation}
Using $\langle \Psi|a^\dagger_\theta a_\theta|\Psi\rangle = 0$ and $\langle \Psi|a_{-\theta}a^\dagger_{-\theta} |\Psi\rangle = \frac L{2\pi}$, we get 
\begin{equation}
    \langle \Psi|b^\dagger_\theta b_\theta|\Psi\rangle = \frac{L}{2\pi} |g_\theta|^2.
\end{equation}
But also, in a GGE with density matrix $\rho_w$, see Sec.~\ref{ssect:freefermion}, we have $\langle b^\dagger_\theta b_\theta\rangle = \frac{L}{2\pi}n(\theta)$, thus we identify
\begin{equation}\label{eq:ng}
    n(\theta) = |g_\theta|^2.
\end{equation}
Again using Eqs.~\eqref{eq:fgconstraints}, we obtain
\begin{equation}
    \frac1{|f_\theta|^2} = \frac1{|g_\theta|^2}
    \frac{|g_{-\theta}|^2}{1-|g_{-\theta}|^2}
\end{equation}
and therefore, from \eqref{eq:occupation} and \eqref{eq:ng},
\begin{equation} \label{eq:Kwapp}
    |\mathcal K_{\theta,-\theta}|^2
    = |\mathcal K_{-\theta,\theta}|^2
    = \frac{|g_{-\theta}|^2}{|f_{-\theta}|^2}
    = \frac{|g_{\theta}|^2}{1-|g_{\theta}|^2}
    = e^{-w(\theta)}.
\end{equation}

Note that in our analysis of GGEs, we assume that $n(\theta)$, thus $|g(\theta)|^2$, has an analytic extension to a neighbourhood of $\mathbb R$. This analyticity property is not true of the function $w(\theta)$, which must have a singularity (e.g.~logarithmic) at $\theta=0$ because of Eq.~\eqref{eq:g0}. We also assume that $n(\theta)\to0$ as $|\theta|\to0$, and thus this must also be true for $g(\theta)$.

Below we report for completeness all the relevant elementary correlation functions of fermionic operators after the quench, and their symmetries. Those will be used in the following subsections, where evaluating current and density correlations, which are bilinears in the fermions (so application of Wick theorem requires only the knowledge of those).  In real space we define
\begin{align} 
    G^{b^\dag b}_{xy}(t,s) &= \bra{\Psi}b^\dag(x,t)b(y,s)\ket{\Psi} 
    \nonumber
    \\
    G^{b^\dag b^\dag}_{xy}(t,s) &= \bra{\Psi}b^\dag(x,t)b^\dag(y,s)\ket{\Psi} .\label{eq:fundamental_correlators}
\end{align}
and similarly for their hermitian conjugates.
Going to momentum space, these take the form
\begin{align}
    &G^{b^\dag b}_{\theta\theta'}(t,s) = \bra{\Psi}b^\dag_\theta(t) b_{\theta'}(s) \ket{\Psi} = e^{\ii E(\theta) (t-s) }|g_\theta|^2  \delta(\theta-\theta')
    \\
    &G^{b^\dag b^\dag}_{\theta\theta'}(t,s) = \bra{\Psi} b^\dag_\theta (t) b^\dag_{\theta'} (s) \ket{\Psi} =- e^{\ii E(\theta) (t+s)}f_\theta g^*_\theta   \delta(\theta+\theta')
    \\
    &G^{b b}_{\theta\theta'}(t,s) = \bra{\Psi} b_\theta(t) b_{\theta'}(s) \ket{\Psi} =- e^{-\ii E(\theta)(t+s) }g_\theta f^*_\theta  \delta(\theta+\theta')
    \\
    &G^{b b^\dag}_{\theta \theta'}(t,s) = \bra{\Psi} b_\theta (t) b^\dag_{\theta'} (s) \ket{\Psi} = e^{-\ii E(\theta) (t-s) }|f_\theta|^2  \delta(\theta-\theta') 
    \label{eq:correlations_table}
\end{align}
where in particular $|g(\theta)|^2 = n(\theta) = 1- |f(\theta)|^2$.
Note the following symmetries 
\begin{align}
    &G^{b^\dag b}_{\theta\theta'}(t,s) = \delta(\theta-\theta')e^{\ii E(\theta)(t-s)} - G^{b b^\dag}_{\theta'\theta}(s,t) 
    \\
    &(G^{b^\dag b}_{xy}(t,s))^* = G^{b^\dag b}_{yx}(s,t) 
    \\
    &G^{b^\dag b}_{xy}(t,s) = \int\frac{d\theta}{2\pi}e^{-\ii \theta(x-y)+\ii E(\theta)(t-s)} - G^{b b^\dag}_{yx}(s,t)  
\end{align}
so that at equal times
\begin{equation}
    G^{b^\dag b}_{xy}(t,t) = \delta_{xy} - G^{b b^\dag}_{yx}(t,t) = \delta_{xy} - (G^{b b^\dag}_{xy}(t,t))^*
    \label{eq:equal_time_bdagb}
\end{equation}
and also
\begin{equation}
    (G^{b^\dag b^\dag}_{\theta \theta'}(t,s))^* = G^{b b}_{\theta'\theta}(s,t)  ,\quad (G^{b^\dag b^\dag}_{xy}(t,s))^* = G^{b b}_{yx}(s,t) .
\end{equation}

\subsection{Approach to the GGE}\label{app:approachGGE}

In the previous Sec.~we have evaluated the GGE $\rho_w$ corresponding to the initial state $|\Psi\rangle$ simply by evaluating the averages of the mode occupation. Here we analyse a bit more in detail how the GGE is approached in time.

We first note that
\begin{equation}
    \bra{\Psi}b^\dagger_\theta b_{\theta'}\ket{\Psi} =   
    \braket{b^\dagger_\theta b_{\theta'}}_{\rho_w}
    = \delta(\theta-\theta')n(\theta).
\end{equation}
Thus, by Wick's theorem, the only difference between averages in $\ket\Psi$ and in $\braket{\cdot}_{\rho_w}$ come from the contraction
\begin{equation}
    \bra\Psi b_\theta b_{\theta'}\ket\Psi
\end{equation}
and its complex conjugate. Thus we evaluate $\bra\Psi b(x,t)b(x',t')\ket\Psi$ in three main situations that are important for our analysis: $t=t',\; x\neq x'$ (for the cumulants of space-integrated conserved densities), and $x=x',\;t\neq t'$ (for the cumulants of time-integrated currents) and $x\neq x', t\neq t'$ (for analysing the correlation between the spatially separated time-integrated currents).

In the first case, we have, using \eqref{eq:babogo},\eqref{eq:fgconstraints}, and the definition in \eqref{eq:fundamental_correlators}
\begin{equation}
G^{b b}_{x x'}(t,t)=
    -\int \frac{d\theta}{2\pi}
    e^{\ii (x -x')\theta -2\ii tE(\theta)}
    f^*_{-\theta}g_{-\theta}.
\end{equation}
Consider $t\to\infty$ with $x,x'$ fixed. Then there is a stationary phase at $\theta_*:E'(\theta_*)=0$, with a resulting integral $\propto \frac1{\sqrt t}$. Thus, this decays as $t\to\infty$: for every two-point functions on intervals that stay finite, the GGE is approached. We notice that as $g_{-\theta_*}=0$ (Eq.~\eqref{eq:g0}),  for fermion two-point functions, the approach is proportional to $1/t^{3/2}$ instead of $1/\sqrt t$; and for multilinears of fermions, the approach is faster.

But we are interested in the scaling $x,x',t\propto \ell\to\infty,\;(x-x')/t=\xi$, for which the exponential has a stationary phase at $\theta_* =\theta_*(\xi):\;E'(\theta_*) = \xi/2$, with a resulting integral $\propto 1/\sqrt\ell$. In charge-neutral fermion bilinears, such as those involved in densities and currents, two such contractions will be multiplied with each other. Thus we have, for instance,
\begin{equation}
    \bra\Psi q(x,t)q(x',t)\ket\Psi^c
    = \braket{q(x,t)q(x',t)}^c_{\rho_w} +
    C(\xi) \,(t\ell)^{-1}
    + O(\ell^{-2}),
\end{equation}
thus the correction is  $O(1/\ell)$. Then, for the cumulant we have
\begin{eqnarray}
    \bra\Psi \int_0^{\ell X} dx\,q(x,\ell t)\,
    \int_0^{\ell X} dx'\,q(x',\ell t)\ket\Psi^c
    &=& \ell^2
    \bra\Psi \int_0^X dx\,q(\ell x,\ell t)\,
    \int_0^X dx'\,q(\ell x',\ell t)\ket\Psi^c
    \nonumber\\
    &\sim& 
    \langle\int_0^{\ell X} dx\,q(x,\ell t)\,
    \int_0^{\ell X} dx'\,q(x',\ell t)\rangle^c_{\rho_w}
    + O(\ell)\nonumber
\end{eqnarray}
where the correction $O(\ell)$ is $\ell \int_{-X/t}^{X/t} d\xi\,(X-2\xi t)C(\xi)$. Therefore, the correction due to the quench changes the linearly scaling part of the cumulant, hence modifies the scaled cumulant from its GGE value (recall that the scaled cumulant is obtained by dividing by $\ell X$, and taking the large $\ell$ limit). Here it would be possible to evaluate explicitly this modification, however it is not necessary for our calculation. The modification due to the quench comes from pair productions -- this will be made much clearer when we study the single-mode densities and currents below.

In fact, there is one limit where it is useful to evaluate this correction term: the limit $X/t\to0$ of $\ell X$-scaled spatially-integrated densities as above. The result for the correction is explicitly
\begin{equation}
    \lim_{X/t\to0}
    \frac{t}{X}\int_{-X/t}^{X/t} d\xi\,\Big(\frac{X}t - 2\xi\Big)C(\xi) = 0
\end{equation}
as $C(\xi)$ is bounded. Thus, in this limit we recover the GGE result. This is in agrement with taking first the long-time limit of the finite-interval cumulant, then the limit of the scaled cumulant on a long interval (this means that the limit $X/t\to0$ is in fact uniform in $t$).

In the second case, where we can set $x=x'=0$, we find, with $E'(\theta_*)=0$ and a saddle point analysis (again remeber the definitions in \eqref{eq:fundamental_correlators}),
\begin{equation}
G^{b b}_{0 0}(t, t')=
    -\int \frac{d\theta}{2\pi}
    e^{\ii (t+t')E(\theta)}
    f^*_{-\theta}g_{-\theta}
    \sim \frac{\sqrt{\ii}\,e^{\ii (t+t')E(\theta_*)}f_{-\theta_*}^*g_{-\theta_*}}{\sqrt{2\pi (t+t')}}.
\end{equation}
As $E(\theta)$ is symmetric, this is $\theta_*=0$, and then, by Eq.~\eqref{eq:g0}, the result vanishes. Therefore, 
\begin{equation}
G^{b b}_{0 0}(t, t')=
    O\Big(\frac1{(t+t')^{3/2}}\Big).
\end{equation}
Hence, the corrections to cumulants of charge-neutral bilinears involve 
\begin{equation}
    \int_1^T dt\,\int_1^Tdt'\, \frac1{(t+t')^2} = O\left(\frac{1}{T^3}\right)
    \ll T \qquad (T\to\infty)
\end{equation}
(where the lower boundary does not matter for the large-$T$ analysis). This correction is sublinear, therefore the quench does not affect cumulants of equal-position time-integrated quantities: for these, the GGE is reached quickly enough. The lack of modification due to the quench comes from the lack of pairs of particles produced at equal (zero) momenta, due to the fermionic statistics.

We remark that if there were particles created at zero momenta (for instance, for bosonic systems), then, still by a calculation similar to that of Eqs.~\eqref{eq:EEk}-\eqref{eq:jjT}, the correction due to the quench would vanish for {\em cumulants of total currents}, which are in any case the objects of interest. Therefore, the fact that pairs of particles of zero momenta are not produced, is not an essential aspects of our calculation.

Finally, we may also analyse time-integrated currents at two different points in a similar way as above, finding:
\begin{eqnarray}
    \bra\Psi \int_0^{\ell T} dt\,j(\ell x, t)\,
    \int_0^{\ell T} dt'\,j(\ell x', t')\ket\Psi^c
    &\sim& 
    \langle
    \int_0^{\ell T} dt\,j(\ell x, t)\,
    \int_0^{\ell T} dt'\,j(\ell x', t')
    \rangle^c_{\rho_w}
    + O(\ell)\ .\nonumber
\end{eqnarray}
This case is necessary for the discussion in Sec.~\ref{ssect:renyisingle}. With $\xi = (x-x')/(t+t')$, the saddle point leading to the $O(\ell)$ correction is at $\theta_*: E'(\theta_*) = \xi$. Thus, the correction due to the quench again changes the linearly scaling part of the two-point cumulant. Here, the limit $\xi\to\infty$ is interesting, and easy to evaluate: as $\xi\to\infty$, the saddle point will be at $\theta_*\to\infty$, and we only have to use the fact that $g_\theta\to0$ as $|\theta|\to\infty$. Therefore, the correction vanishes as $\xi\to\infty$, and we may use the GGE result, where scaled cumulants of time-integrated currents become sums of cumulants at $x$ and at $x'$ in the GGE (which take the same values by translation invariance).

\subsection{Single-mode density and currents and decay of correlations in GGEs}\label{app:singlemode}

In the main text, when studying the dynamics of the entanglement of an interval after a quench (Sec. \ref{ssect:renyisingle}), we are interested in the single-mode twist fields. This requires constructing densities and currents not only for the global $U(1)$ charge as done above, but also for the individual conserved quantities $b^\dagger_\theta b_\theta$. These conserved quantities are not extensive -- they are not integrals of local or quasi-local observables -- however, as reviewed in \cite{DeNardis_2022}  in the more general context of integrable models, they form a scattering basis for such extensive quantities. Thus, integrations over small $\theta$-intervals give extensive conserved quantities. These are the single-mode conserved quantities that we now investigate.

As we are working directly in the thermodynamic limit and, in the continuum of space, the momenta fill the real axis $[-\infty,\infty]$. Let us write this as a union of disjoint intervals centered at equispaced "target" momenta: $\cup_{i=-\infty}^{\infty}A_{\theta_i}$ where $A_\theta=[\theta-\epsilon/2, \theta+\epsilon/2)$ and $\theta_{i} = (i+1/2)\epsilon$. We can write the total charge as
\begin{equation}
    Q = \int dx\, b^\dag(x) b(x) = \int d\theta b^\dag_\theta b_\theta =\sum_{i=-\infty}^{\infty}Q_{\theta_i},\qquad 
    Q_{\theta} = \int_{A_{\theta}}d\theta'b^\dag_{\theta'} b_{\theta'}.
    \label{eq:appendixA:single_mode_charge}
\end{equation}
Clearly each ``regularised" (by $\epsilon$) single-mode charge $Q_\theta$ is conserved,  $\commutator{Q_\theta}{H}=0$. It is also extensive: in a GGE in a finite volume $L$, we have $\braket{Q_\theta^2}^c\propto L$:
\begin{equation}
    \braket{Q_\theta^2}^c
    = \int_{\theta-\epsilon/2}^{\theta+\epsilon/2} d\theta' d\theta''\,
    \delta(\theta'-\theta'')^2n(\theta')(1-n(\theta'))
    = \frac L{2\pi}
    \int_{\theta-\epsilon/2}^{\theta+\epsilon/2} d\theta'
    n(\theta')(1-n(\theta'))\ .
\end{equation}

As mentioned, if we want to write a density in real space for each $b^\dag_{\theta}b_{\theta}$, we will get something non local. However, $Q_{\theta}$'s have quasi-local densities. We seek a function $f_{\theta}(x,y)$ such that 
\begin{equation}
    \int dx dy\, b^\dag(x) b(y) f_{\theta}(x-y) =Q_{\theta}.
\end{equation}
Going to Fourier space, one can show that (see \eqref{bnormalisation})
\begin{equation}
    f_{\theta}(z) =  \frac{\sin(\frac{\epsilon z}{2} )}{\pi z}e^{\ii \theta z} .
\end{equation}
The corresponding regularised single-mode density,  parametrised by the momentum, and one choice of the density (the only hermitian and PT symmetric one), is given by
\begin{equation}
    q_\theta(x,t) = \int dz\, b^\dag(x+z/2,t)b(x-z/2,t)f_\theta(z). \label{eq:appendixA:single_mode_density}
\end{equation}
In terms of Fourier modes, this takes the form
\begin{equation}
    q_\theta(x,t) = \int \frac{dk dk'}{2\pi}\, 
    e^{\ii x(k'-k)} \vartheta\left(\frac{\epsilon}2 - \Big|\frac{k+k'}2-\theta\Big|\right)b^\dag_k(t) b_{k'}(t).
\label{eq:appendixA:single_mode_density_Fourier}
\end{equation}
As $\commutator{Q_\theta}{H}=0$, the density $q_\theta(x,t)$ has an associated current satisfying a continuity equation and by a calculation analogous to \eqref{eq:appendixA:local_current} one finds
\begin{equation}
    j_\theta(x,t) = \int\frac{dk dk'}{2\pi}e^{\ii x(k'-k)}\left(\frac{E(k')-E(k)}{k'-k}\right)\vartheta\left(\frac{\epsilon}2 - \Big|\frac{k+k'}2-\theta\Big|\right)b^\dag_k(t) b_{k'}(t)\label{eq:appendixA:single_mode_current}
\end{equation}
where $\vartheta(x)$ is the Heaviside theta function. This is basically the same as \eqref{eq:appendixA:local_current} with a restriction on the $q$ integration around the target mode $\theta$.

For convenience, in fact we will consider the momentum-pair densities and currents
\begin{equation}\label{eq:pairmodedensities}
    q_{|\theta|}(x,t) = q_\theta(x,t) + q_{-\theta}(x,t),\quad
    j_{|\theta|}(x,t) = j_\theta(x,t) + j_{-\theta}(x,t)
\end{equation}
associated to a pair of opposite momenta; these are the ones used in the twist field decomposition \eqref{eq:taufactsingletwopoint}.

We now analyse the behavior of two-point functions on GGEs following the same lines of \ref{app:global}. This is because, again, the original BFT was defined for expectation values on GGEs and, eventually, we would like to replace with those the expectations on the initial state before the quench. Later, we will study more carefully the approach to the GGE.
In a GGE (characterized by Boltzmann factor $w(k)$), the only surviving elementary correlator is
\begin{equation}
    \braket{b^\dag_k(t) b_{k'}(s)} = e^{\ii E(k) (t-s)} \delta(k-k') n(k)
\end{equation}
with $n(k) = (1+e^{w(k)})^{-1}$.

We now consider the connected correlation functions of densities and currents along the paths of interest and study under which condition they decay fast enough in such GGE.
The two paths of interest are the horizontal path $(0,t)\to(x,t)$, and the piecewise linear path $(0,t)\to(0,0)\to(x,0)\to(x,t)$ (cf. Fig. \ref{quench_finitesystem} in the main text).
For the first case, we need to evaluate density-density correlations at equal time and different spatial points.
In the second case, instead, we need to evaluate current-current correlations both between the two different vertical segments and within the segments themselves (using the same arguments of Sec. \ref{ssect:renyihalf}, one can show that the contribution of the horizontal segment vanishes, so we do not need to look at density-currents correlations as well).
The calculation is slightly different than in the case of \ref{app:global} because we have to deal with pair-mode densities and currents. The global quantities are determined directly by the correlators $\braket{b^\dag(x,t) b(y,s)}$, while here we have to analyse directly the density-density or current-current correlations. 

Let us start by considering the connected density-density correlation function on the horizontal path. We focus on single-mode quantities, the pair-mode ones being just linear combination of those, i.e.,
\begin{align}
    \braket{q_{|\theta|}(x,t)q_{|\theta|}(0,t)}^c &= \braket{q_\theta(x,t)q_\theta(0,t)}^c + \braket{q_\theta(x,t)q_{-\theta}(0,t)}^c 
    \nonumber\\
    &+ \braket{q_{-\theta}(x,t)q_\theta(0,t)}^c + \braket{q_{-\theta}(x,t)q_{-\theta}(0,t)}^c \label{eq:appendixA:theta_correlator_density}
\end{align}
Below, in the current subsection, all expectations $\langle \cdot \rangle$ are on the GGE. Let $\eta = \pm 1$, we have
\begin{align} \label{eq:qth-qth_GGE}
    &\braket{q_{\theta}(x,t)q_{\eta\theta}(x',t)}^c
    \nonumber
    \\
    &=\int \frac{dk dk'}{2\pi}\frac{dq dq'}{2\pi}\, 
    e^{\ii x(k'-k)} \vartheta\left(\frac{\epsilon}2 - \Big|\frac{k+k'}2-\theta\Big|\right)
    \nonumber
    \\
    &\times e^{\ii x'(q'-q)} \vartheta\left(\frac{\epsilon}{2}-\left|\frac{q+q'}{2}-\eta\theta\right|\right)\braket{b^\dag_k(t) b_{k'}(t) 
    b^\dag_q(t) b_{q'}(t)}^c
    \nonumber
    \\
    & = \int \frac{dk dk'}{2\pi}\frac{dq dq'}{2\pi}\, 
    e^{\ii x(k'-k)} \vartheta\left(\frac{\epsilon}2 - \Big|\frac{k+k'}2-\theta\Big|\right)
    \nonumber
    \\
    &\times e^{\ii x'(q'-q)} \vartheta\left(\frac{\epsilon}2 - \Big|\frac{q+q'}2-\eta\theta\Big|\right)\braket{b^\dag_k(t)b_{q'}(t)}\braket{b_{k'}(t)b_{q}^\dag(t)}
    \nonumber
    \\
    &= \int \frac{dk dk'}{2\pi}e^{\ii (x-x')(k'-k)}\vartheta\left(\frac{\epsilon}2 - \Big|\frac{k+k'}2-\theta\Big|\right)\vartheta\left(\frac{\epsilon}2 - \Big|\frac{k+k'}2-\eta\theta\Big|\right)(1-n(k'))n(k)\quad. 
\end{align}
Note that all four correlators needed can be deduced from the this upon exploiting $\theta\mapsto-\theta$ transformation.
Apart from the overall step function which constrains one of the integrals, with the assumptiong on $n(k)$ that we have (it is analytic on a neighbourhood of the real line, and decays rapidly enough as $|k|\to\infty$), we can change variables to $k_\pm = k'\pm k$ and perform the $k_-$ integral in the complex plane by letting $k_-\mapsto k_- - \ii \gamma \sign(x-x')$ (and $\gamma>0$) and we see that the decay is exponential making cumulants scale as in \eqref{eq:cumulant_scaling}. Therefore, BFT is applicable.

For the current-current correlator, we look at two generic points in space and time. However, by making use of time-translational invariance of the GGE, we can set one time to zero. Analogous manipulations give (we focus on $x>0$)
\begin{align} \label{jxt-j0s_0}
    \braket{j_{\theta}(x,t)j_{\eta\theta}(0,0)}^c &= \int \frac{dk dk'}{2\pi}e^{\ii x(k'-k) + \ii t(E(k)-  E(k'))}\vartheta\left(\frac{\epsilon}2 - \Big|\frac{k+k'}2-\eta\theta\Big|\right)
    \nonumber
    \\
    &\times\vartheta\left(\frac{\epsilon}2 - \Big|\frac{k+k'}2-\theta\Big|\right)\left(\frac{E(k')-E(k)}{k'-k}\right)^2(1-n(k'))n(k)
\end{align}
and we see that that in the ballistic scaling limit, with $  \zeta = x/t $ fixed, the exponential has a saddle point at $\partial_k E(k_*) = \partial_{k'}E(k_*') = \zeta$ so that for $\zeta \neq 0, +\infty$ (recall eq. \eqref{eq:EEk} and the one below)
\begin{align} \label{jxt-j0s}
    \braket{j_{\theta}(x,t)j_{\eta\theta}(0,s)}^c &\sim \frac{e^{-\ii (t-s) E(k_*(\zeta))}}{2\pi (t -s)}(\partial_k E(k_*(\zeta)))^2\vartheta\left(\frac{\epsilon}2 - \Big|k_*(\zeta)-\theta\Big|\right)\vartheta\left(\frac{\epsilon}2 - \Big|k_*(\zeta)-\eta\theta\Big|\right)
    \nonumber\\
    &\times(1-n(k_*(\zeta)))n(k_*(\zeta)) \quad .
\end{align}

Let us take the case $\eta=1$ : the two step functions square to one and the condition for it to vanish is $k_*(\zeta)\geq \epsilon/2+ \theta$ or $k_*(\zeta)\leq -\epsilon/2+ \theta$. Using $\zeta = \partial_k E(k_*(\zeta)) = v(k_*(\zeta))$ and the monotonicity of the velocity, we can invert the above relation, leading to a vanishing step functions whenever 
\begin{equation}
    \zeta \geq v(\theta+\epsilon/2)=v(\theta) + O(\epsilon) \quad \textbf{or}\quad  \zeta \leq v(\theta-\epsilon/2)= v(\theta)- O(\epsilon)
    \label{condition-jxt0s}
\end{equation}
For $\eta=-1$ we obtain the same condition because we are considering $\zeta \geq 0$.  
This means that, under the condition \eqref{condition-jxt0s}, the leading term of correlation \eqref{jxt-j0s} (coming from the saddle point) vanishes.

Correlations between the two vertical segments, on the contrary, arise only in a tiny cone (of order $\epsilon$) around the velocity $v(\theta)$. When those are present, the second cumulant resulting from \eqref{jxt-j0s} grows as $O(T \log T)$ at large time $T$ (to be contrasted with \eqref{eq:jjT}).


Note that if we consider the special case $t=s$ ( $\zeta=+\infty$), then \eqref{jxt-j0s} has no saddle point and therefore its decay is exponentially fast.

In the above calculation we assumed $x\neq 0$ ($\zeta\neq0$), which corresponds at looking at correlations between the two different aforementioned vertical paths (see again Fig. \ref{quench_finitesystem}, main text). In order to study the decay of current-current correlations along any of such vertical paths, instead, we need to set equal space (and different times). Again, using translational invariance we just set $x=0$ in \eqref{jxt-j0s_0}.
In this case we can repeat the argument in \eqref{eq:EEk} and below to show
\begin{equation} \label{j00-j0t}
       \braket{j_{|\theta|}(0,t)j_{|\theta|}(0,0)}^c = O\big(\frac1{t^3}\Big)\qquad (t\to\infty)
\end{equation}
with the corresponding cumulant scaling linearly in time.

Finally, we recall again that the horizontal segment of the path 
$(0,t) \to (0,0) \to (x,0)\to (x,t)$
does not contribute to the cumulants. This observation, together with \eqref{j00-j0t}, and when the condition \eqref{condition-jxt0s} is satisfied, allows to conclude that within a GGE: (i) the SCGF associated to the two-point function of the pair-mode twist fields can be correctly evaluated along this path  (i.e., BFT is applicable); (ii) the total SCGF factorises into those associated to the two vertical cuts, namely it is the sum of the two corresponding SCGFs (again we refer to Remark 3.3 in \cite{doyon2020fluctuations}), and, in fact, in the main text, we make use of such factorization.

\subsection{Approach to the GGE}\label{ssect:decay}

Finally in this subsection, following \ref{app:approachGGE}, we study in more detail the approach in time to the corresponding GGE values of the same single-mode densities and currents correlations considered in \ref{app:singlemode}.
In particular we want to understand how cumulants in the GGE's are modified at large time when taking into account correlations coming from the initial state $|\psi\rangle$ (cf. Eq. \eqref{squeezed}). Below, we denote by $\langle \cdot \rangle$ expectation values on such initial state, while $\langle \cdot \rangle_{\rho_{w}}$ the ones on the GGE atteined at infinite time.

We start with the single-mode connected density-density (again the corresponding pair-mode correlator is obtained via \eqref{eq:appendixA:theta_correlator_density}).
In momentum space we have (we focus on $x>0$)
\begin{align}
    \braket{q_{\theta}(x, t)q_{\eta\theta}(0,t)}^c &= \int\frac{dkdk'}{2\pi}\frac{dqdq'}{2\pi}e^{ix(k'-k)}\vartheta\left(\frac{\epsilon}{2}-\left|\frac{k+k'}{2}-\theta\right|\right)\vartheta\left(\frac{\epsilon}{2}-\left|\frac{q+q'}{2}-\eta\theta\right|\right)
    \nonumber\\
    &\times\braket{b^\dag_k(t)b_{k'}(t)b^\dag_q(t)b_{q'}(t)}^c
    \nonumber
    \\
    & = \int\frac{dkdk'}{2\pi}\frac{dqdq'}{2\pi}e^{ix(k'-k)}\vartheta\left(\frac{\epsilon}{2}-\left|\frac{k+k'}{2}-\theta\right|\right)\vartheta\left(\frac{\epsilon}{2}-\left|\frac{q+q'}{2}-\eta\theta\right|\right)
    \nonumber
    \\
    &\times\left(\braket{b^\dag_k(t)b_{q'}(t)}\braket{b_{k'}(t)b_{q}^\dag(t)} - \braket{b^\dag_k(t)b_{q}^\dag(t)}\braket{b_{k'}(t)b_{q'}(t)}\right)
\end{align}
where $\eta=\pm1$.
The first piece is nothing but the GGE contribution and, as discussed in \ref{app:singlemode} (see Eq. \eqref{eq:qth-qth_GGE} and below) is always well-behaved in the ballistic regime.

The other part gives
\begin{align} \label{corrections_qth-qth}
     \int\frac{dkdk'}{(2\pi)^2}&e^{ix(k'-k) +2iE(k) t - 2i E(k')t}\vartheta\left(\frac{\epsilon}{2}-\left|\frac{k+k'}{2}-\theta\right|\right)\vartheta\left(\frac{\epsilon}{2}-\left|\frac{k+k'}{2}-\eta\theta\right|\right) (1-n(k))n(k')
\end{align}
In the ballistic limit, with $\zeta = x/t$ fixed, we can use again saddle point analysis, similarly to calculations in \ref{app:singlemode}. 
Now, however, the saddle points is given by $ 2v(k') - \zeta = 0$ and $2v(k) = \zeta$ (note the factor 2 of difference wrt the saddle point of the integral \eqref{jxt-j0s_0}), namely
$k^*(\zeta) = k'^*(\zeta) = v^{-1}(\zeta/2)$. After application of saddle-point method, the $\vartheta$ function is evaluated at these points. For $\eta=1$, using $\vartheta^2=\vartheta$, the result of the saddle point of \eqref{corrections_qth-qth} vanishes unless $\left| \frac{k^*(\zeta)+k'^*(\zeta)}{2} -\theta \right|  \le \frac{\epsilon}{2}$  or, equivalently, $| v^{-1}(\zeta/2) - \theta | \le \epsilon/2$ that it precisely
\begin{equation} \label{condition_qthqth_quench}
    v(\theta- \epsilon/2)\leq \zeta/2 \leq v(\theta+ \epsilon/2) \ .
\end{equation}
Having assumed $x$>0, for $\eta = -1$ we recover the very same condition.

This is exactly the expected condition for two particles of opposite momentum $\pm \theta$, initially forming an entangled pair, not to hit both the segment $[0,x]$ at time $t$ (as be easily understood geometrically from Fig.~\ref{quench_finitesystem} of the main text).
Note that, for $\zeta$ fixed (and due to the above mentioned factor 2), this condition is more severe then Eq. \eqref{condition-jxt0s}, so it also guarantees a fast enough decay of same correlation within the GGE.

When condition \eqref{condition_qthqth_quench} does not hold, from the saddle point contribution, we get that the integral \eqref{corrections_qth-qth} decays as $t^{-1}$, giving a slow approach to the GGE. This, in fact, modifies the behaviour of the second cumulant because it gives a correction to the GGE value which is not subleading (it is actually faster than linear in time). Outside the range \eqref{condition_qthqth_quench} the correction to the GGE decays fast enough, so that the associated cumulant is not modified at leading order.



Let us briefly comment the situation in the case of correlation functions of the single-mode current even though the main idea exactly follows the density-density case. The relevant quantity in this case is 
\begin{align}
    \braket{j_{\theta}(x,t)j_{\eta\theta}(0, s)}^c&= \int\frac{dkdk'}{2\pi}\frac{dqdq'}{2\pi}e^{ix(k'-k)}\vartheta\left(\frac{\epsilon}{2}-\left|\frac{k+k'}{2}-\theta\right|\right)\vartheta\left(\frac{\epsilon}{2}-\left|\frac{q+q'}{2}-\eta\theta\right|\right)
    \nonumber
    \\
    &\times\left(\frac{E(k')-E(k)}{k'-k}\right)^2\left(\braket{b^\dag_k(t)b_{q'}(s)}\braket{b_{k'}(t)b_{q}^\dag(s)} - \braket{b^\dag_k(t)b_{q}^\dag(s)}\braket{b_{k'}(t)b_{q'}(s)}\right)\label{eq:appendixA:theta_correlator_current}\quad .
\end{align}
This time we cannot use time-translation invariance as done before when computing the expectation on a GGE.
The first part in the above expression is again the GGE contribution alayzed in \eqref{jxt-j0s}, while the second comes from quasi-particle pairs and gives rise to saddle points in the ballistic regime. 
It can be checked that the condition for the saddle point contribution not to vanish is exactly the same as for the density, Eq.\eqref{condition_qthqth_quench}. 
%
%
Therefore, only when the condition \eqref{condition_qthqth_quench} does not hold, the contribution of the second term in \eqref{eq:appendixA:theta_correlator_current} is subleading wrt to the GGE one, and the associated cumulant is not shifted with respect to the leading GGE value.

Therefore, depending on the value of $\theta$ (apart for $\theta$ in a region of order $\epsilon$, which is to trace back to our regularization of the observables) either the correlations of single-mode densities or those of single-mode currents show a fast approach to their GGE value, thus imposing the right path to choose when using BFT.



\section{$S$ matrix in the $\alpha-$copy theory}
\label{app:Salpha_matrix}

Consider $S(\theta,\theta')$ to be the $S$ matrix in the single-copy
theory (we consider the diagonal case for simplicity, but everything
below can be generalized to non-diagonal $S$ matrices), i.e. 
\begin{equation}
|\theta,\theta'\rangle=S(\theta,\theta')|\theta',\theta\rangle
\end{equation}
(namely, it is the factor we get by exchanging 
$\theta,\theta'$ in the two-particle state), then one can define
in the $\alpha-$copies theory
\begin{equation}
S^{(\alpha)}(\theta,i;\theta',i')=\delta_{{\rm mod}(\alpha)}^{ii'}S(\theta,\theta')\pm(1-\delta_{{\rm mod}(\alpha)}^{ii'})\label{eq:Salpha_matrix}
\end{equation}
acting on the state $|\theta,i;\theta',i'\rangle$, where we introduced explicitly the dependence on the copy-index.
Importantly, the $\pm$ sign in (\ref{eq:Salpha_matrix}) depends
on the commutation relations of the fields among the copies. 

Note that Eq. \eqref{eq:Salpha_matrix} is the S-matrix associated to $\alpha$ \emph{independent} copies, which only describes the symmetry of the $\alpha$-copies theory (i.e., it does not take into account the constraints of the fields in different copies implemented by the twist fields exchange relations (cf. \eqref{eq:permexch1}-\eqref{eq:permexch2})). However this is enough for our purposes.

By going to Fourier space in the replica index $\mathcal{F}_{i\to p}$,
we have
\begin{equation}
|\theta,p;\theta',p'\rangle=\sum_{p,p'}S^{(\alpha)}(\theta,\theta';p,p';k,k')|\theta',k';\theta,k\rangle
\end{equation}
with (by simple algebra)
\begin{equation}
S^{(\alpha)}(\theta,\theta';p,p';k,k')=\delta_{{\rm mod}(\alpha)}(k+k'-p-p')\left[S(\theta,\theta')\pm1\right]\mp\delta_{{\rm mod}(\alpha)}(p-k)\delta_{{\rm mod(}\alpha)}(p'-k')\:.
\end{equation}

This can be written explicitly as a $2\alpha\times2\alpha$ matrix
$S^{(\alpha)}(\theta,\theta')_{m,n}$ for $\alpha\in\mathbb{N}$, with row and column indices
 $m=\left\{ p,p'\right\} $ and $ n=\left\{ k,k'\right\}$ respectively. For example, for $\alpha=2$, it takes
the form 
\begin{equation}
S^{(2)}(\theta,\theta')=\begin{bmatrix}S(\theta,\theta') & 0 & 0 & S(\theta,\theta')\pm1\\
0 & S(\theta,\theta') & S(\theta,\theta')\pm1 & 0\\
0 & S(\theta,\theta')\pm1 & S(\theta,\theta') & 0\\
S(\theta,\theta')\pm1 & 0 & 0 & S(\theta,\theta')
\end{bmatrix}\label{eq:S2_matrix}
\end{equation}
Note that one can check that, as expected, this $S$ matrix satisfy
the Yang-Baxter equations for general $S(\theta,\theta')$.

Now, for the specific case of free fermions, consider Eq. (\ref{eq:fourier_transform_replica})
in the main text. In the first basis, the fields $\psi_{i}$ commutes
among different copies: in this case above we will choose the $+$
sign in $S^{(\alpha)}$ (cfr. Eq. (\ref{eq:Salpha_matrix})). However,
after the $SU(\alpha)$ trasformation to the $\psi_{j}$ basis, the
fields in different copies also anticommute, and this amout to choosing
the sign $-$ in (\ref{eq:Salpha_matrix}). Since for free fermions
we have $S(\theta,\theta')=-1$, then we see that in the $\psi_{j}$
basis, $S^{(\alpha)}$ becomes diagonal (cfr. Eq.~(\ref{eq:S2_matrix})).

\end{appendix}



\bibliography{biblio_renyi.bib}

\nolinenumbers

\end{document}